\documentclass[12pt]{spieman}  
\pdfoutput=1
\hypersetup{pdfauthor={Montgomery et al.},pdftitle={Performance of the SPT-3G DfMUX readout}}
\usepackage{tocloft}
\usepackage[sort, numbers]{natbib}
\usepackage{amsmath,amsfonts,amssymb}
\usepackage{graphicx}
\usepackage{authblk}
\makeatletter
\def\@author{}
\renewcommand\@author{\ifx\AB@affillist\AB@empty\AB@author\else
	\ifnum\value{affil}>\value{Maxaffil}\def\rlap##1{##1}%
	\AB@authlist \\[\affilsep]\AB@affillist
	\else  \AB@authors\fi\fi}
\makeatother
\usepackage{pdflscape}
\usepackage{subcaption}
\usepackage[colorlinks=true, allcolors=blue]{hyperref}
\usepackage{tikz}
\usepackage[american,siunitx]{circuitikzgit}
\usetikzlibrary{arrows.meta, positioning}
\usepackage{booktabs, siunitx}
\DeclareSIUnit{\sqrthz}{\ensuremath{\sqrt{\text{\hertz}}}}
\DeclareSIUnit{\parthz}{\pico\ampere/\sqrt{\hertz}}
\DeclareSIUnit{\ukarcmin}{\micro\kelvin\text{-arcmin}}
\DeclareSIUnit{\sqdeg}{\text{deg\ensuremath{^{2}}}}
\DeclareSIUnit{\deg}{\text{deg}}
\DeclareSIUnit{\arcmin}{\text{arcmin}}
\defcitealias{dobbs12b}{D12}

\newcommand{\sptg}{\textsc{SPT-3G}}

\newcommand{\litebird}{LiteBIRD}
\newcommand{\sptpol}{SPTpol}

\newcommand{\order}{\mbox{${\cal O}$}}

\newcommand{\parallelsum}{\mathbin{\|}}

\newcommand{\rev}[1]{{\color{black}#1}}

\usepackage{lineno}
\title{Performance and characterization of the SPT-3G digital frequency-domain multiplexed readout system using an improved noise and crosstalk model}

\author[a]{J.~Montgomery}
\author[b]{P.~A.~R.~Ade}
\author[c,d]{Z.~Ahmed}
\author[e]{E.~Anderes}
\author[f,g]{A.~J.~Anderson}
\author[h]{M.~Archipley}
\author[i]{J.~S.~Avva}
\author[j]{K.~Aylor}
\author[k]{L.~Balkenhol}
\author[l,g]{P.~S.~Barry}
% ...

\affil[a]{Department of Physics and McGill Space Institute, McGill University, 3600 Rue University, Montreal, Quebec H3A 2T8, Canada}
\affil[b]{School of Physics and Astronomy, Cardiff University, Cardiff CF24 3YB, United Kingdom}
\affil[c]{Kavli Institute for Particle Astrophysics and Cosmology, Stanford University, 452 Lomita Mall, Stanford, CA, 94305, USA}
\affil[d]{SLAC National Accelerator Laboratory, 2575 Sand Hill Road, Menlo Park, CA, 94025, USA}
\affil[e]{Department of Statistics, University of California, One Shields Avenue, Davis, CA 95616, USA}
\affil[f]{Fermi National Accelerator Laboratory, MS209, P.O. Box 500, Batavia, IL, 60510, USA}
\affil[g]{Kavli Institute for Cosmological Physics, University of Chicago, 5640 South Ellis Avenue, Chicago, IL, 60637, USA}
\affil[h]{Department of Astronomy, University of Illinois at Urbana-Champaign, 1002 West Green Street, Urbana, IL, 61801, USA}
\affil[i]{Department of Physics, University of California, Berkeley, CA, 94720, USA}
\affil[j]{Department of Physics \& Astronomy, University of California, One Shields Avenue, Davis, CA 95616, USA}
\affil[k]{School of Physics, University of Melbourne, Parkville, VIC 3010, Australia}
\affil[l]{High-Energy Physics Division, Argonne National Laboratory, 9700 South Cass Avenue., Argonne, IL, 60439, USA}
% ...

\cftpagenumbersoff{figure}
\cftpagenumbersoff{table} 
\begin{document} 
\maketitle

\begin{abstract}

The third generation South Pole Telescope camera (\sptg{}) improves upon its predecessor (\sptpol{}) by an order of magnitude increase in detectors on the focal plane. The technology used to read out and control these detectors, digital frequency-domain multiplexing (DfMUX), is conceptually the same as used for \sptpol{}, but extended to accommodate more detectors. 
A nearly 5x expansion in the readout operating bandwidth has enabled the use of this large focal plane, and \sptg{} performance meets the forecasting targets relevant to its science objectives. However, the electrical dynamics of the higher-bandwidth readout differ from predictions based on models of the \sptpol{} system \rev{due to the higher frequencies used, and parasitic impedances associated with new cryogenic electronic architecture.} To address this, we present an updated derivation for electrical crosstalk in higher-bandwidth DfMUX systems, and identify two previously uncharacterized contributions to readout noise, \rev{which become dominant at high bias frequency.} The updated crosstalk and noise models successfully describe the measured crosstalk and readout noise performance of \sptg{}. \rev{These results also suggest specific changes to warm electronics component values, wire-harness properties, and SQUID parameters, to improve} the readout system for future experiments using DfMUX, such as the \litebird{} space telescope.
\end{abstract}


{\noindent \footnotesize\textbf{*} Joshua Montgomery,  \linkable{Joshua.J.Montgomery@mcgillcosmology.ca} }

\begin{spacing}{1}   

\section{Introduction}
\label{sec:intro}  
The South Pole Telescope (SPT) is a 10-meter telescope located at the geographic South Pole as part of the Amundsen--Scott Research Station. SPT is used to observe the sky at microwave frequencies, with the goal of making deep and high-resolution measurements of the cosmic microwave background (CMB); it is currently equipped with the \sptg{} receiver, the third camera to be deployed on the telescope. \sptg{} is in its third year of surveying a 1,500\,deg$^2$ field of the CMB, using a polarization-sensitive tri-chroic focal plane with 16,000 detectors \cite{anderson18}. The detectors are bolometric transition-edge sensors (TES): metal films held at sub-Kelvin temperatures in the transition between normal and superconducting states \cite{irwin05}. TES detectors convert depositions of incident power to variations in the resistance of the film; these variations are sensed by applying a voltage bias across the TES and measuring the resulting current through the circuit. This methodology is sufficiently sensitive to detect \order(\SI{10}{\atto\watt}) fluctuations in deposited power.
TES devices have been the standard for the past three generations of SPT receivers, and are common throughout the field of CMB instrumentation, in part because they operate at or near the photon noise limit. For this reason, each generation of receiver for SPT has improved sensitivity primarily by increasing the number of TES devices operated simultaneously, and \sptg{} observes the sky with an order of magnitude more detectors than \sptpol{}, decommissioned in 2017.

One of the enabling technologies for increasing focal plane size is the multiplexed readout, which allows multiple TES detectors to be operated with a shared set of electronics.
Without multiplexing, detector numbers would be constrained by cryogenic cooling limitations and the cost of the readout system. As detector numbers increase, improvements to multiplexing technology are necessary. 
A multiplexing readout system includes both room-temperature signal processing electronics and cryogenic analog electronics, and is characterized by the number of detectors that can be operated as a single module of shared electronics (the multiplexing factor or ``mux factor''). \sptg{} employs a 68x mux design that is conceptually based on the 16x mux system of \sptpol{}, but extended to accommodate higher multiplexing.

Electrical models used to design and forecast the performance of \sptg{} were derived from the 16x mux predecessors, and make a number of approximations or assumptions that are no longer valid in the higher mux factor regime. 
Consequently, achieved electrical crosstalk and readout noise performance are worse than expected. 
Despite this, performance remains either within target requirements (in the case of crosstalk), or sufficient with respect to target scientific analysis (in the case of elevated readout noise).
The performance difference from expectation indicates an incomplete understanding of the system dynamics, which we seek to correct with the model updates presented here.
In Section \ref{sec:dfmux} we give a description of the readout system that highlights relevant non-idealities. In Section \ref{sec:crosstalk} we derive updated analytic forms for the crosstalk in such systems. In Section \ref{sec:noise} we present the improved noise model alongside measurements from \sptg{}, as well as descriptions of two new mechanisms relevant to accurately modeling \sptg{} readout noise. \rev{These mechanisms stem from a previously un-modeled output filter (Section \ref{sec:output_filter_tf}) and a capacitive current path that generates feedback-mediated noise (Section \ref{sec:current_sharing_tf}).}
Results from this updated set of models are being used to inform specific design choices to improve existing instruments using this technology and future instruments such as the \litebird{} space telescope.

\section{Digital Frequency-Domain Multiplexing} \label{sec:dfmux}

The need to multiplex is dictated by the requirement that TES devices be kept at sub-Kelvin temperatures ($\sim$\SI{270}{\milli\kelvin} in the case of \sptg{}). 
Without multiplexing, each detector would be connected to room temperature by a separate pair of conductors, producing a total heat load in excess of the cooling power available.
\sptg{} overcomes this limitation by using a multiplexing strategy known as frequency-domain multiplexing (FDM). The particular FDM designs used on the SPT cameras are known as fMUX, starting with the SPT-SZ camera in 2007, which used an analog frequency-domain multiplexing (AfMUX) system \cite{dobbs12b}. This was superseded by the digital frequency-domain multiplexing (DfMUX) system deployed on the \sptpol{} instrument in 2011 \cite{dobbs08}.
\sptg{} uses the second generation DfMUX readout, which was first introduced in \cite{bender19}. The description given here will omit details not relevant for the crosstalk and noise model updates, but a detailed account of the modern DfMUX design, and full noise modeling, can be found in \cite{montgomery20}. 

\begin{figure}[hbtp]
	\centering
	\includegraphics[width=0.9\textwidth]{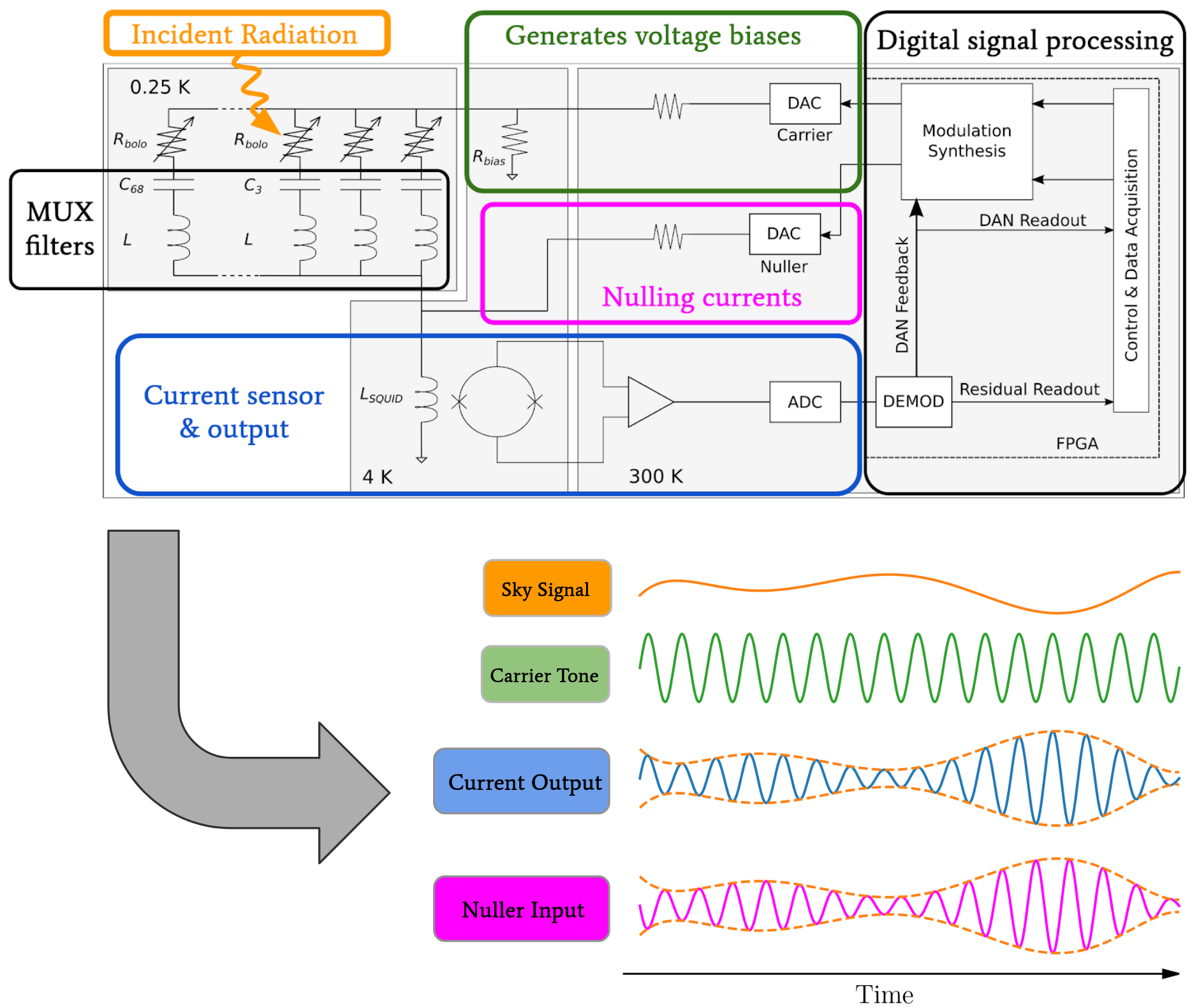} 
	\captionsetup{width=0.9\linewidth} 
	\vspace*{0.5cm}
	\caption[DfMUX schematic (simplified)]{A simplified schematic diagram of the DfMUX readout system. Voltage biases are calculated and produced as a composite waveform in the warm electronics; these are divided into component sinusoids by a bank of cryogenic filters; incident radiation deposits power on the TES detectors, changing their resistance; this amplitude-modulates the carrier sinusoids and produces a current waveform with the sky signal encoded in the sidebands, similar to AM radio; the current waveforms are summed together into another composite waveform that is sensed in the output signal path. To linearize the SQUID amplifiers, a nulling waveform is generated that cancels signals at the input of the SQUID using active feedback. The nuller waveform is then used as the science data output, since it accurately reproduces the sky signal in order to cancel the current waveform. The lower inset figure shows example sky, carrier, output, and nuller signals for a single TES. Figure adapted from \cite{bender18}.}
	\label{fig:dfmux_simple}
\end{figure}

DfMUX preserves the independence of each TES bias, while limiting the number of cryogenic wires required, by applying the bias voltages as megahertz sinusoids (the \textit{carriers}). 
In the previous generation of DfMUX readout up to 16 of these carriers were distributed at frequencies between \SI{200}{\kilo\hertz} and \SI{1.2}{\mega\hertz}, but current designs operate up to 68 such carriers in a bandwidth up to \SI{5.5}{\mega\hertz}.
The individual carrier tones are summed together in room-temperature electronics to generate a composite waveform that can be transmitted to the sub-Kelvin stage over a single pair of conductors.
That waveform is separated back into the component sinusoids at the sub-Kelvin stage, using a bank of cryogenic resonant filters \cite{bender16}.
Carrier frequencies are chosen to correspond to filter resonant frequencies, and each TES is embedded within a filter. 
This allows a bias at the proper frequency to be applied to a TES, while isolating it from bias voltages intended for other TES devices.
As the TES detectors vary in resistance they amplitude-modulate the associated carrier tone, generating a current waveform in which the sky signal is encoded. Each of these amplitude-modulated tones is then summed to make a single output waveform.
The output waveform is sensed cryogenically using a superconducting quantum interference device (SQUID), before being amplified by conventional electronics and digitally demodulated to recover the independent sky signals incident on each TES.
This operation is analogous to AM radio, and is shown schematically in Figure \ref{fig:dfmux_simple}.

\subsection{Nulling}  \label{sec:nulling}

\begin{figure}[hbtp]
	\centering
	\includegraphics[width=0.7\textwidth]{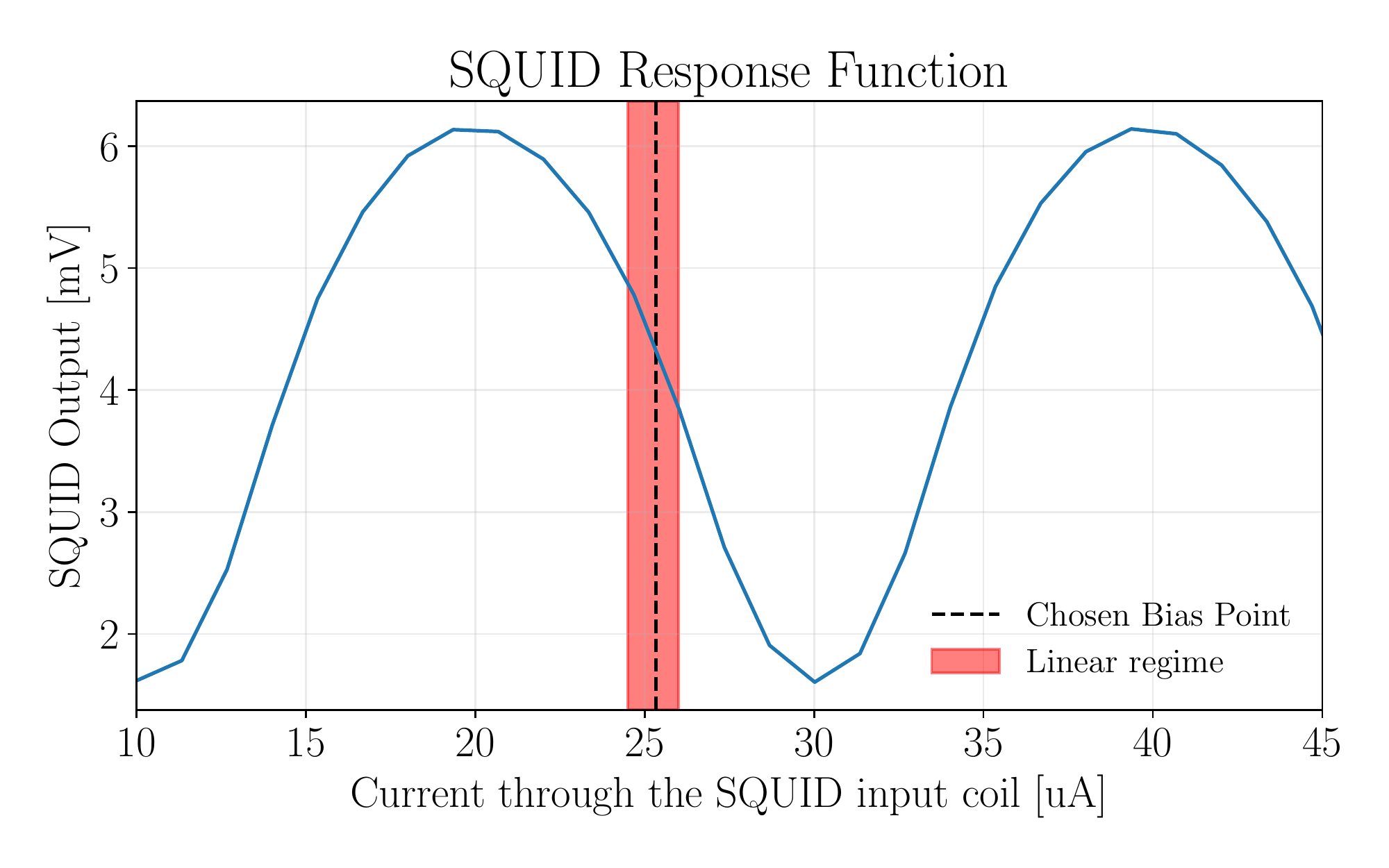} 
	\captionsetup{width=0.7\linewidth} 
	\caption[SQUID response function.]{The SQUID output response is a periodic function of the input current, resulting in limited useful dynamic range. Annotated is the approximately linear response regime and the bias point at which we operate the SQUIDs. \rev{The regime indicated visually is a heuristic, based on the change of SQUID performance as a function of dynamic range used.} The bias point refers to a DC input bias (the ``flux bias'')  used to center the SQUID response in the linear regime. Figure from \cite{montgomery20}.}
	\label{fig:squid_nonlinear_response}
\end{figure}

SQUID devices act as transimpedance amplifiers, converting current at the input to a voltage at the output. They are highly non-linear, with a periodic response function and limited dynamic range (Figure \ref{fig:squid_nonlinear_response}).
SPT-3G uses arrays of individual SQUIDs configured in parallel and series banks and fabricated by the National Institute of Standards and Technology \cite{doriese15, stiehl11, bender18}. A typical SPT-3G SQUID array will exhibit significant non-linearity for inputs greater than $\sim$\SI{2.1}{\micro\ampere}, which is less than the current produced by a single TES voltage bias.
To linearize the SQUID amplifiers a separate current waveform is injected at the SQUID input to cancel the incoming signals. That current waveform is called a ``nuller,'' and is generated using narrow-band digital feedback centered at the carrier bias frequencies. The bandwidth of the feedback is sufficient to capture the science signals in the sidebands of the carrier tones; under this scheme our data is the signal generated by the feedback, rather than the output signal from the SQUID.
This feedback system is known as Digital Active Nulling (DAN) and was first described in \cite{dehaan12} for the first-generation DfMUX readout system. An account of DAN for modern higher-density systems can be found in \cite{montgomery20}. 

\subsection{Parasitic capacitances to ground}

The schematic readout diagram shown in Figure \ref{fig:dfmux_simple} lacks stray circuit elements of the system that are relevant at the higher bias frequencies now used. The most notable of these are capacitances to ground within the cryogenic filter elements.
The filters are made up of 2D lithographic devices that generate capacitance and inductance using geometric shapes. The shapes have trace widths between \SI{4}{\micro\meter} and \SI{16}{\micro\meter} and total areas of up to \SI{\sim20}{\square\milli\meter} \cite{hattori14,rotermund16,montgomery20}.
A byproduct of this design is a parallel-plate capacitance with the ground plane \SI{675}{\micro\meter} away. 
There are similar capacitances to ground throughout the readout electronics, formed by the microstrip routing on the TES wafer and traces of the separate PCBs on which the filters and SQUIDs are mounted.
All of these may be estimated based on the design of the lithography or layout of the electronics cards. 

\begin{figure}[h]
	\centering
	\includegraphics[width=\linewidth]{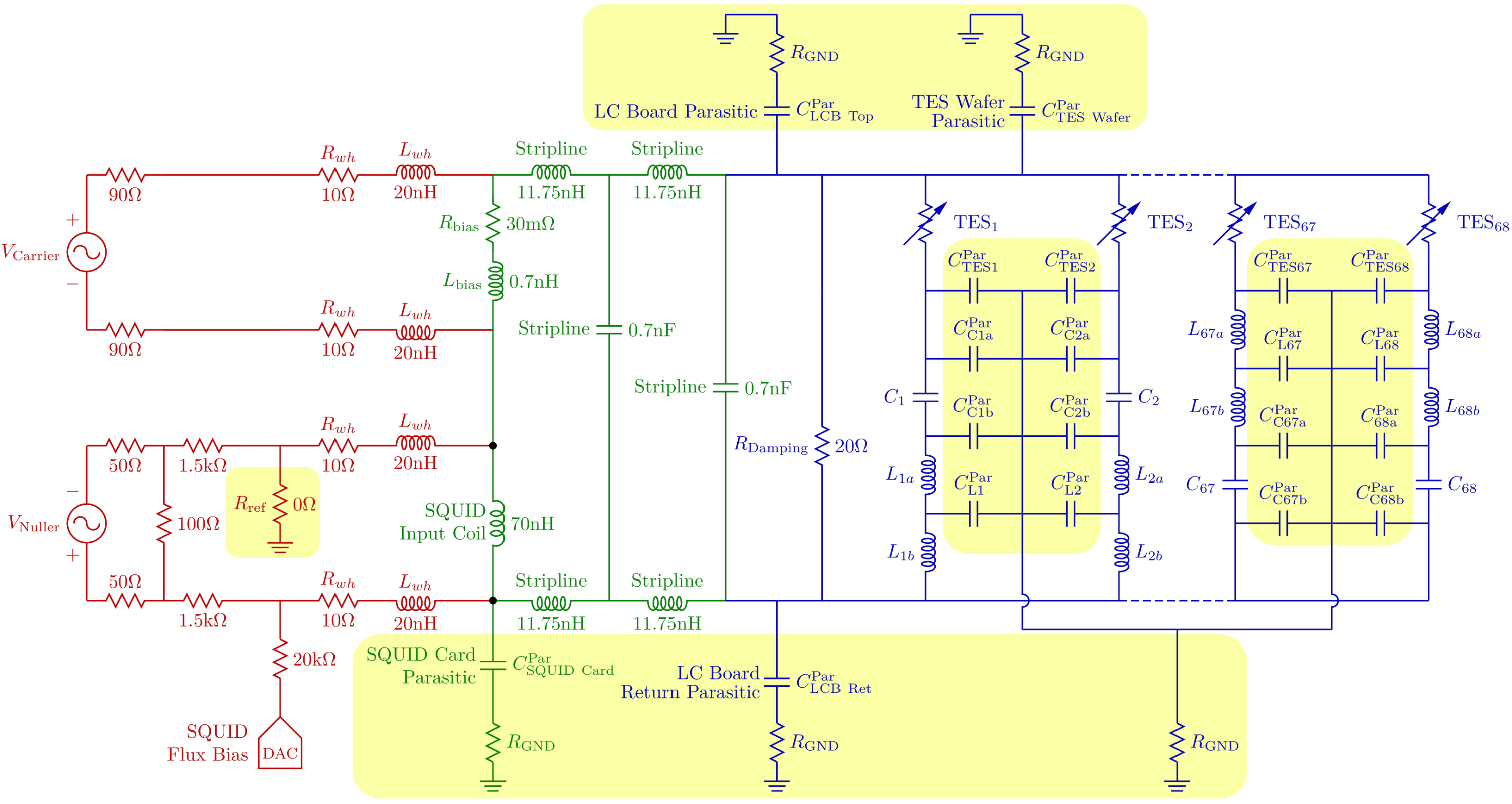}
	\captionsetup{width=0.95\linewidth}
	\vspace*{1mm}
	\caption[Electrical circuit model schematic.]{An electrical model of the signal chain that includes stray impedances, such as an inductance in series with the bias resistor \cite{elleflot20}, series inductance in the striplines \cite{avva18}, and parasitic capacitances to ground \cite{montgomery20,dutcher18}. Red corresponds to electronics at room temperature through to the wire harness. Green corresponds to electronics at the 4K stage through to striplines that connect the 4K and sub-Kelvin stages. Blue indicates electronics at the sub-Kelvin stage that include the LC filtering board and TES wafer. Parasitic strays new to this circuit model and relevant to the discussions herein are highlighted in yellow. $R_\mathrm{GND}$ is typically low impedance. Channels 1 \& 2 demonstrate the RCL configuration, while channels 67 \& 68 demonstrate an alternative RLC configuration that is also used for \sptg{} filters. Not pictured are the initial amplification or filtering stages common to both the carrier and nuller outputs, or the output path between the SQUID output and the ADC.}
	\label{fig:circuit_model_complete}
\end{figure}

Capacitances to ground are largely irrelevant at lower bias frequencies, but at higher frequencies they present a low enough impedance path through the system to modify transfer functions. This effect was first noted in \cite{dutcher18} with respect to how stray capacitances bias measurements of detector properties. We show in Section \ref{sec:current_sharing_tf} that these parasitic current paths can also significantly amplify readout noise.
A more complete circuit model for the synthesizer signal path is given in Figure \ref{fig:circuit_model_complete}.

\section{Alternative forms of base-band feedback}
\rev{Digital active nulling is one of several forms of base-band feedback (BBFB); the first of which was developed by the SRON group \cite{denhartog09}, and is implemented differently. 
Where DAN linearizes the SQUID by providing feedback to a summing junction shared by the carrier voltages (as shown in Figure \ref{fig:dfmux_simple}), the SRON BBFB implementation does so using an independent SQUID feedback coil that is not directly coupled to the carrier voltage or the input coil \cite{sakai14}.
Similarities between the two systems allow efforts to understand systematics or perform modeling for one system to be generalized to the other.
This is not always the case though, as the differences have driven divergent design choices that can produce (or eliminate) some sources of systematics.
This work covers mechanisms that fall into both of these categories, and so the applicability will be summarized briefly.}

\subsection{Crosstalk and BBFB}

\rev{The relevant three crosstalk mechanisms for both systems are the same (leakage current, leakage power, and inductor-inductor magnetic coupling), although the relative importance of each form can differ substantially. The basic forms of these mechanisms were introduced in \cite{vanderkuur04}, and analytic equations introduced in \citep[][hereafter D12]{dobbs12b} are used to estimate crosstalk for both systems \cite{hijmering14, bender19}; including, most recently, a review of electrical crosstalk in \cite{wang21}.

In this work, we show that the analytical models from \citetalias{dobbs12b} misestimate leakage current and leakage power crosstalk due to a series of built-in assumptions or simplifications.
We then update the \citetalias{dobbs12b} models and validate these improvements using \sptg{} data.
Alternative means of calculating crosstalk, such as from numerical simulations of the full circuit \cite{denhartog18} are independent to the forms of these models.
}

\subsection{Feedback-mediated noise and BBFB}

\rev{In SRON-like BFBB systems the TES detectors are in series with the common impedance of the input coil to the SQUID, whereas for a DAN system the active feedback at the summing junction across that input coil makes a virtual ground, eliminating this source of impedance.
This enables DAN systems to use relatively high-input-inductance single-stage SQUID amplifiers, while SRON systems use ultra-low-input-inductance SQUIDs, most recently in a two-stage system \cite{kiviranta2018}.

A disadvantage of DAN feedback is feedback-mediated readout noise. Since the SQUID input inductance can be large, and it is shared with the carrier circuit, there exist two competing paths for current through the feedback circuit. This results in a magnification of readout noise in the presence of DAN feedback, an effect called ``current sharing,'' first reported in \cite{bender18}. 
In SRON-like BBFB designs, the separate feedback coil to the SQUID is only weakly coupled to the carrier circuit via transformer coupling with the input coil. This makes any competing current paths in the feedback circuit higher impedance, such that the design is likely not susceptible to this mechanism of readout-noise enhancement; with \cite{hijmering16} specifically reporting no change in readout noise in the presence of base-band feedback.

The work in this paper builds on \cite{bender18} by identifying a third important current path for \sptg{}.
This time through parasitic capacitances in the readout circuit, which couple back through a ground reference in the warm electronics used to bias the single-stage SQUID.
Unlike the current sharing described in \cite{bender18}, which is fundamental to DAN-like architectures, this mechanism is particular to the specific implementation of the DAN readout system and cryogenic electronics currently in use on \sptg{}.}

\subsection{SQUID output-filter mediated noise}

\rev{
Single-stage SQUIDs used in DAN-like systems typically have a large output impedance, followed by a wiring harness to room-temperature electronics. This makes the signal path susceptible to low-pass filtering by parallel capacitance in the wiring harness. 
Such a filter effectively increases the readout noise at the attenuated frequencies.
The combination of output impedance and wiring harness capacitance used in \sptg{} attenuates signals in the same frequency range that carrier tones are operated, such that the associated noise increase is present in bolometer data.

This provides a useful point of comparison between different architectures. This filtering in the \sptg{} system shorts out high frequency out-of-band signals, preventing resonances in the SQUID that degrade performance \cite{elleflot20b, montgomery20}. In some SRON systems, this function is intentionally performed by a snubber, which acts as a low-pass filter \cite{audley2020} with a cutoff frequency outside of bolometer carrier frequencies. 
Theoretically, an SRON-like BBFB system could be subject to the form of noise reported here, but the snubber design and low output-impedance SQUIDs make it unlikely, and it has not been reported in SRON system publications.
}

\section{Crosstalk} \label{sec:crosstalk}

Electrical coupling between detectors in a DfMUX system is a dominant source of crosstalk in the instrument. This can occur in two ways:
\begin{description}
	\item[Leakage current crosstalk] happens when the \textit{i\textsuperscript{th}} carrier voltage is amplitude-modulated by variations in the \textit{n\textsuperscript{th}} TES within a different filter. This is caused by overlap between the filter bandwidths, which allows some current to leak through one of the other parallel legs of the cryogenic filter. This effect is a strong function of the filter shape and spacing, and causes signal from the \textit{n\textsuperscript{th}} TES to crosstalk into the output from the \textit{i\textsuperscript{th}} TES. 
	\item[Leakage power crosstalk] happens because leakage current deposits electrical power across the TES in the legs it leaks through. Under some conditions, deposited power across the \textit{n\textsuperscript{th}} TES due to leakage current can vary as a function of the \textit{i\textsuperscript{th}} TES resistance. This varying leakage power mimics the power depositions from the sky, and causes signal from the \textit{i\textsuperscript{th}} TES to crosstalk into the output from the \textit{n\textsuperscript{th}} TES.
\end{description}
These mechanisms were first derived for an fMUX system in \citetalias{dobbs12b}, and those derivations were used to model the expected crosstalk performance of the \sptg{} design.
While the mean crosstalk performance met our design requirement \cite{bender19}, the phenomenology diverges noticeably from the expectation. 
This is largely a consequence of additional stray impedances relevant to crosstalk, which were not characterized or included in the \citetalias{dobbs12b} derivations. A more complete derivation of these crosstalk mechanisms is given in Section \ref{sec:xtalk_derivation}. The resulting total crosstalk is described in Section \ref{sec:xtalk_fraction}, and Section \ref{sec:xtalk_validation} uses in-situ crosstalk measurements to validate the model updates.
The primary differences with the \citetalias{dobbs12b} model are:
\begin{enumerate}
	\item Inclusion of stray series resistance within each parallel leg of the LC filter network. This effect is the most prominent source of differences between the two models for the SPT-3G design.
	\item Preservation of phase information. The \citetalias{dobbs12b} model approximates total crosstalk fraction using the magnitudes of each of the crosstalk and primary signals. However, these signals can be out-of-phase with one another, leading to cancellation and suppression. This effect is relatively small for an \sptg{}-like design, but can be significant for systems with large series impedance with the cryogenic filters.
	\item Bias frequency flexibility. The \citetalias{dobbs12b} model assumes bias frequencies are exactly at the filter resonance. In higher-bandwidth systems, it is common for bias frequencies to be offset from the true resonant frequency by up to a few hundred hertz, inducing extra complex impedance. These offsets are due to fluctuations in resonant frequencies as a function of TES resistance, and how bias frequencies are chosen to mitigate inter-modulation distortion products \cite{montgomery20}. \rev{This model decouples the chosen bias frequency from the associated resonance frequency.}
\end{enumerate}

\subsection{Crosstalk model derivation} \label{sec:xtalk_derivation}

\begin{figure}[hbtp]
	\centering
	\begin{circuitikz}[scale=0.75]
		
		\draw (0,-0.5) to [short] (0,-5)
		to [sV,v=$V_{\mathrm{bias}}$, l_=Carrier] (0,-8)
		to [short] (0,-14);
		\draw (0,-0.5) to [short] (10,-0.5)
		to [short] (10,-2);
		\draw  (7,-2) to [short] (10.5,-2); 
		\draw[thick, dotted] (10.5,-2) to (12.5,-2);
		\draw  (12.5,-2) to [short] (13,-2); 
		
		\draw (7,-2) to [short] (7,-2.5) 
		to [vR, l=$R_{\mathrm{TES,1}}$] (7,-5)
		to [european resistor, l=$r_{s,1}$] (7,-7)
		to [L, l=$L_1$] (7,-9)
		to [C, l=$C_1$] (7,-11)
		to [short] (7,-11);
		\draw (10,-2) to [short] (10,-2.5) 
		to [vR, l=$R_{\mathrm{TES,2}}$] (10,-5)
		to [european resistor, l=$r_{s,2}$] (10,-7)
		to [L, l=$L_2$] (10,-9)
		to [C, l=$C_2$] (10,-11);
		
		\draw (13,-2) to [short] (13,-2.5) 
		to [vR, l=$R_{\mathrm{TES,68}}$] (13,-5)
		to [european resistor, l=$r_{s,68}$] (13,-7)
		to [L, l=$L_{68}$] (13,-9)
		to [C, l=$C_{68}$] (13,-11)
		to [short] (13,-11);
		
		\draw (7,-11) to [short] (10.5,-11);
		\draw[thick, dotted] (10.5,-11) to (12.5,-11);
		\draw (12.5,-11) to [short] (13,-11);		
		
		\draw (10,-11) to [short] (10,-11.5)
		to [european resistor, l=$Z_{\mathrm{com}}$] (10,-13.75)
		to [short] (10,-14)
		to [short] (0,-14);
		
		\draw[thick, dashed, red] (5.5,-1.25)   to (15.75, -1.25);
		\draw[thick, dashed, red] (15.75, -1.25) to (15.75, -11.5);
		\draw[thick, dashed, red] (15.75, -11.5) to (5.5,-11.5);
		\draw[thick, dashed, red] (5.5,-11.5) to (5.5,-1.25);
		\node[text=red] at (12,-0.75) {$Z_\mathrm{net}$};
		
	\end{circuitikz}
	\captionsetup{width=0.9\linewidth}
	\vspace*{0.45cm}
	\caption[Cryogenic filter circuit.]{An example circuit diagram of the cryogenic network. This includes all relevant components used in the derivation of leakage current crosstalk and leakage power crosstalk.}
	\label{fig:series_impedance_model_circuit}
\end{figure}
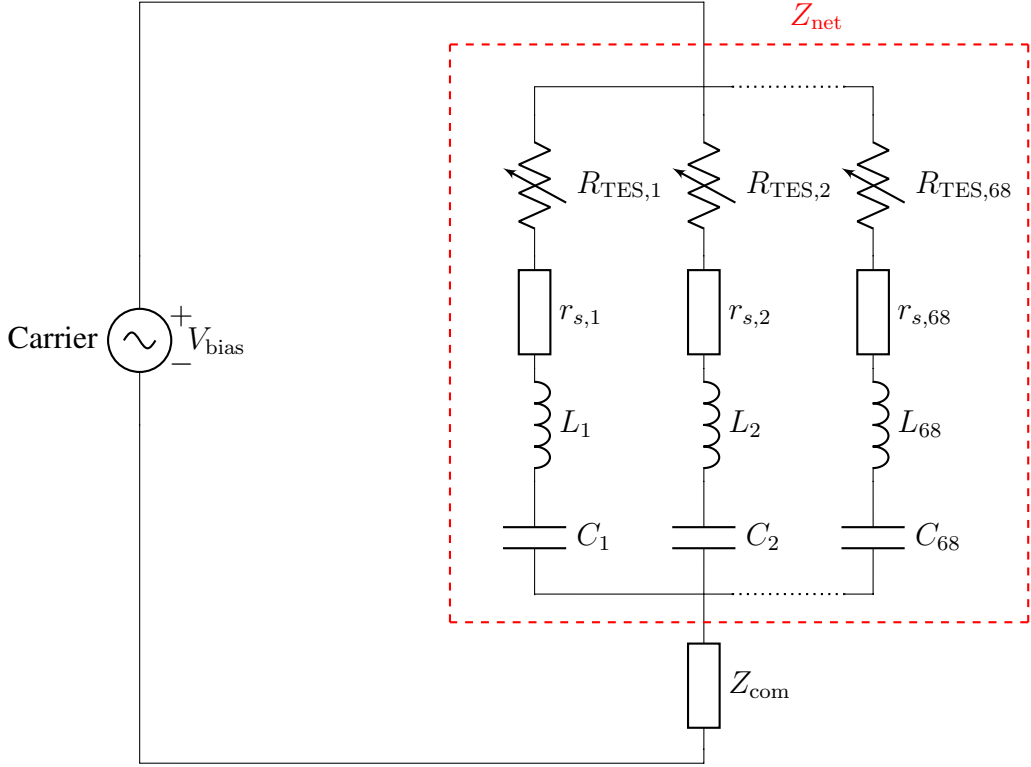

The simplified circuit model shown in Figure \ref{fig:series_impedance_model_circuit} captures the relevant electrical elements for calculating crosstalk: 
\begin{enumerate}
	\item A voltage source producing carrier sinusoids at frequencies $\omega_{i}$.
	\item The filtering network, formed by a parallel bank of series LCR filters. Each leg of the filter includes both a TES and a stray series resistance.
	\item A common impedance in series with the filtering network ($Z_\mathrm{com}$). In the \sptg{} system this is dominated by the inductive reactance of the SQUID input coil (which is suppressed when DAN is active), and of the cryogenic striplines between 4K and sub-Kelvin stages (which remains, even when DAN is active). 
\end{enumerate}
This circuit model omits the nulling path and the parasitic capacitances.
The nulling path modifies $Z_\mathrm{com}$ to remove the SQUID input impedance in series with the cryogenic filter, which is equivalent to a different choice of $Z_\mathrm{com}$, in our case corresponding to the stripline reactance.
Differences in parasitic capacitance within each leg can modify the relative impedance of the parallel legs, but not enough to meaningfully change crosstalk dynamics. 
Typical \sptg{} parameters for the elements in this circuit model are summarized in Table \ref{table:spt3g_params}.
{\renewcommand{\arraystretch}{1.25}
	\begin{table}[hbtp]
		\begin{center}
			\small
			\begin{tabular}{lll}
				\\
				\multicolumn{2}{c}{\textbf{Typical \sptg{} Parameters}}\\
				Parameter & Value \\
				\toprule
				$Z_\mathrm{com}$ & $j \omega\cdot$(\SI{46}{\nano\henry}) \\
				$R_\mathrm{TES}$ & \SIrange{1.3}{1.7}{\ohm} \\
				$r_\mathrm{s}$ & \SIrange{0.25}{0.4}{\ohm} \\
				$L$ & \SI{60}{\micro\henry} \\
				$C$ & \SIrange{12}{150}{\pico\farad} \\
				$2\pi\omega_i$ & \SIrange{1.6}{5.5}{\mega\hertz}, log-spaced \\
				\bottomrule
				\\
			\end{tabular}
			\captionsetup{width=0.7\linewidth}
			\caption[Typical circuit parameters for \sptg{}.]{Typical \sptg{} circuit parameters relevant to the calculation of electrical crosstalk. Resonant frequencies are designed by varying the capacitance to distribute the frequencies logarithmically within the bandwidth, so the lowest frequency regime has the narrowest frequency spacing. Bias frequencies are selected based on the resonant frequencies exhibited, but may be offset from the exact resonant frequency by up to several hundred hertz.}
			\label{table:spt3g_params}
		\end{center}
\end{table}}

Following Figure \ref{fig:series_impedance_model_circuit}, for bias frequency $\omega_i$, the impedance of any single cryogenic filter leg $n$ is
\begin{equation} \label{eq:zn_params}
Z_{n,i} = R_{\mathrm{TES},n} + r_{s,n} + j\omega_i L_{n} + \frac{1}{j\omega_i C_{n}}\,,
\end{equation}
such that $R_{\mathrm{TES},n}$ is the TES resistance, $r_{s,n}$ is any stray series resistance with the TES, and $L_n$ and $C_n$ are the inductor and capacitor elements that define the filter resonant frequency.
In this notation the impedance of the ``on-resonance'' cryogenic leg is $Z_{i,i}$, while $Z_{n\neq i,i}$ are the impedances of ``off-resonance'' cryogenic legs.
The impedance of the full parallel network is
\begin{equation}
Z_{\mathrm{net}}(\omega_i) = \left(~\sum_{n=1}^\mathrm{mux~factor} \frac{1}{Z_{n,i}}\right)^{-1}\,.
\end{equation}

$i\pm1$ are ``nearest-neighbors'' with respect to $i$, such that $Z_{i\pm 1,i}$ are the two lowest impedance paths through the cryogenic network at $\omega_{i}$, aside from the on-resonance leg.
Significant crosstalk coupling only occurs between on-resonance detectors and nearest-neighbors.
For \sptg{} design parameters, $Z_{i\pm 1,i}$ ranges from \SIrange{20}{80}{\ohm}, while $Z_{i,i}$ are typically $<$\SI{2}{\ohm}. 
Therefore, the impedance of the network at each bias frequency may be approximated
\begin{equation}  \label{eq:off_res_z_approx}
Z_{\mathrm{net}}(\omega_i) \approx Z_{i,i}\,.
\end{equation}

\subsubsection{Primary signal}

The intended primary signal is the change in current at frequency $\omega_i$ due to changes in the on-resonance TES resistance $R_{\mathrm{TES},i}$:

\begin{align}
\left[\frac{\delta I_i}
{\delta R_{\mathrm{TES},i}}\right]_{\mathrm{signal}} &={} \frac{\delta}{\delta R_{\mathrm{TES},i}}
\left[\frac{V_{\mathrm{bias}}(\omega_{i})}
{Z_{i,i}+Z_{\mathrm{com}}(\omega_{i})}\right]\\
\delta I_{i,\mathrm{signal}} &={} \frac{-V_{\mathrm{bias}}(\omega_{i})\cdot \delta R_{\mathrm{TES},i}}
{(Z_{i,i} + Z_{\mathrm{com}}(\omega_{i}))^2} \,.\label{eq:science_signal}
\end{align}
Crosstalk happens when $\left(\dfrac{\delta I_i}{\delta R_{\mathrm{TES},n\neq i}}\right) \neq 0$.

\subsubsection{Leakage current crosstalk}

Leakage current crosstalk occurs because some fraction of the current induced through $Z_\mathrm{net}$ by $V_{\mathrm{bias}}(\omega_{i})$ flows through an off-resonance leg $n\neq i$, allowing variations in both the on-resonance and off-resonance TES to amplitude-modulate $V_{\mathrm{bias}}(\omega_{i})$.
This form of crosstalk is given by
\begin{align}
\left[\frac{\delta I_i}{\delta R_{\mathrm{TES},n\neq i}}\right]_{\mathrm{LC}} ={} &\frac{\delta}{\delta R_{\mathrm{TES},n}}\left[\frac{V_{\mathrm{bias}}(\omega_{i})}{(Z_{n,i}+Z_{\mathrm{com}}(\omega_{i}))}\right]\;.
\end{align}
Expressed as a leakage current signal, $\delta I_{i,n,\mathrm{LC}}$, this becomes
\begin{align}
\delta I_{i,n,\mathrm{LC}}={} &\frac{-V_{\mathrm{bias}}(\omega_{i})\cdot \delta R_{\mathrm{TES},n}}
{(Z_{n,i} + Z_{\mathrm{com}}(\omega_{i}))^2}\,. \label{eq:leakage_current_di}
\end{align}
Or as a crosstalk fraction, using Equation \ref{eq:science_signal},
\begin{align} \label{eq:leakage_current_xtalk}
\left[\frac{\delta I_{i,n,\mathrm{LC}}}{\delta I_{i,\mathrm{signal}}}\right] &=\frac{\delta R_{\mathrm{TES},n}}{\delta R_{\mathrm{TES},i}}
\left(\frac{Z_{i,i} + Z_{\mathrm{com}}(\omega_{i})}
{Z_{n,i} + Z_{\mathrm{com}}(\omega_{i})}\right)^2\,.
\end{align}
For most practical systems in which detectors have approximately uniform resistance and saturation powers, $\delta R_{\mathrm{TES},n}\approx \delta R_{\mathrm{TES},i}$, and the first term drops out. Variations in TES resistance produce signals in a narrow bandwidth ($<$100\si{\hertz}) relative to the filter bandwidths ($>$10\si{\kilo\hertz}), so crosstalk fractions calculated at the bias frequencies are sufficient to describe the crosstalk across all sideband signals of interest. Note that common impedances to the filtering network ($Z_{\mathrm{com}}$) contribute to leakage current crosstalk, but are not required for it.

\subsubsection{Leakage power crosstalk}

Leakage current dissipates power across the TES detector in the off-resonance leg through which it flows; this is called leakage power.
Leakage power deposited onto the \textit{i\textsuperscript{th}} detector from leakage current induced by the \textit{n\textsuperscript{th}} voltage bias is given by
\begin{equation}\label{eq:leakage_power_full}
P_{i,n} = \left( \frac{Z_{\mathrm{net}}(\omega_{n}) V_{\mathrm{bias}}(\omega_{n})}
{Z_{\mathrm{net}}(\omega_{n}) + Z_{\mathrm{com}}(\omega_{n})}\right)^2
\frac{R_{\mathrm{TES},i}}
{Z_{i,n}^2}\,.
\end{equation}
In the simple case when $Z_\mathrm{com}=0$, leakage power deposited across any TES is only a function of that TES resistance, and therefore no crosstalk mechanism exists.
When $Z_{\mathrm{com}} \neq 0$, an additional voltage divider is formed with the cryogenic network. What was a fixed voltage bias across the filtering network $Z_\mathrm{net}$ now varies as a function of $Z_\mathrm{net}(\omega_{n})$.
Variations in $R_{\mathrm{TES},n}$ then modulate $P_{i,n}$, generating a form of crosstalk
\begin{align}\label{eq:leakage_power_didr_der}
\left[\frac{\delta I_i}{\delta R_{\mathrm{TES},n\neq i}}\right]_{\mathrm{LP}} ={}& \frac{1}{V_{\mathrm{bias}}(\omega_{i})}
\left[\frac{\delta P_{i,n}}{\delta R_{\mathrm{TES},n}}\right] \;.
\end{align}
Expressed as a leakage power signal, $\delta I_{i,n,\mathrm{LP}}$, this becomes
\begin{equation}\label{eq:leakage_power_di}
\delta I_{i,n,\mathrm{LP}}\approx{}
\frac{V^{2}_{\mathrm{bias}}(\omega_{n})}
{V_{\mathrm{bias}}(\omega_{i})}
\left(\frac{Z_{n,n} ~ Z_\mathrm{com}(\omega_{n})}
{(Z_{n,n} + Z_{\mathrm{com}}(\omega_{n}))^3}\right)
\left(\frac{2~R_{\mathrm{TES},i}~\delta R_{\mathrm{TES},n}}{Z^2_{i,n}}\right)\,,
\end{equation}
with the approximation from Equation \ref{eq:off_res_z_approx} applied.
The equation in the form of a crosstalk fraction is
\begin{align} \label{eq:leakage_power_xtalk}
\left[\frac{\delta I_{i,n,\mathrm{LP}}}{\delta I_{i,\mathrm{signal}}}\right]
\approx -
\left(	 \frac{V^2_{\mathrm{bias}}(\omega_{n})\delta R_{\mathrm{TES},n}}
{V^2_{\mathrm{bias}}(\omega_{i})\delta R_{\mathrm{TES},i}}
\right)
\frac{(Z_{i,i} + Z_\mathrm{com}(\omega_{i}))^2}
{(Z_{n,n} + Z_{\mathrm{com}}(\omega_{n}))^3}
\left(	\frac{2~R_{\mathrm{TES},i}~Z_{n,n} ~ Z_\mathrm{com}(\omega_{n})}
{Z^2_{i,n}}
\right)\,.
\end{align}
\rev{The first term of this expression drops out under the assumption that detectors are approximately uniform in responsivity and saturation power (and therefore bias voltage).
In practice, TES non-uniformity tends to follow a physical gradient across a wafer, while nearest-neighbors are physically co-located on the wafer. This minimizes the overall sensitivity of the crosstalk to variations in TES parameters.}

\subsection{Total crosstalk fraction} \label{sec:xtalk_fraction}

\begin{figure}[hbtp]
	\centering
	\begin{subfigure}[b]{0.7\textwidth}
		\includegraphics[width=\textwidth]{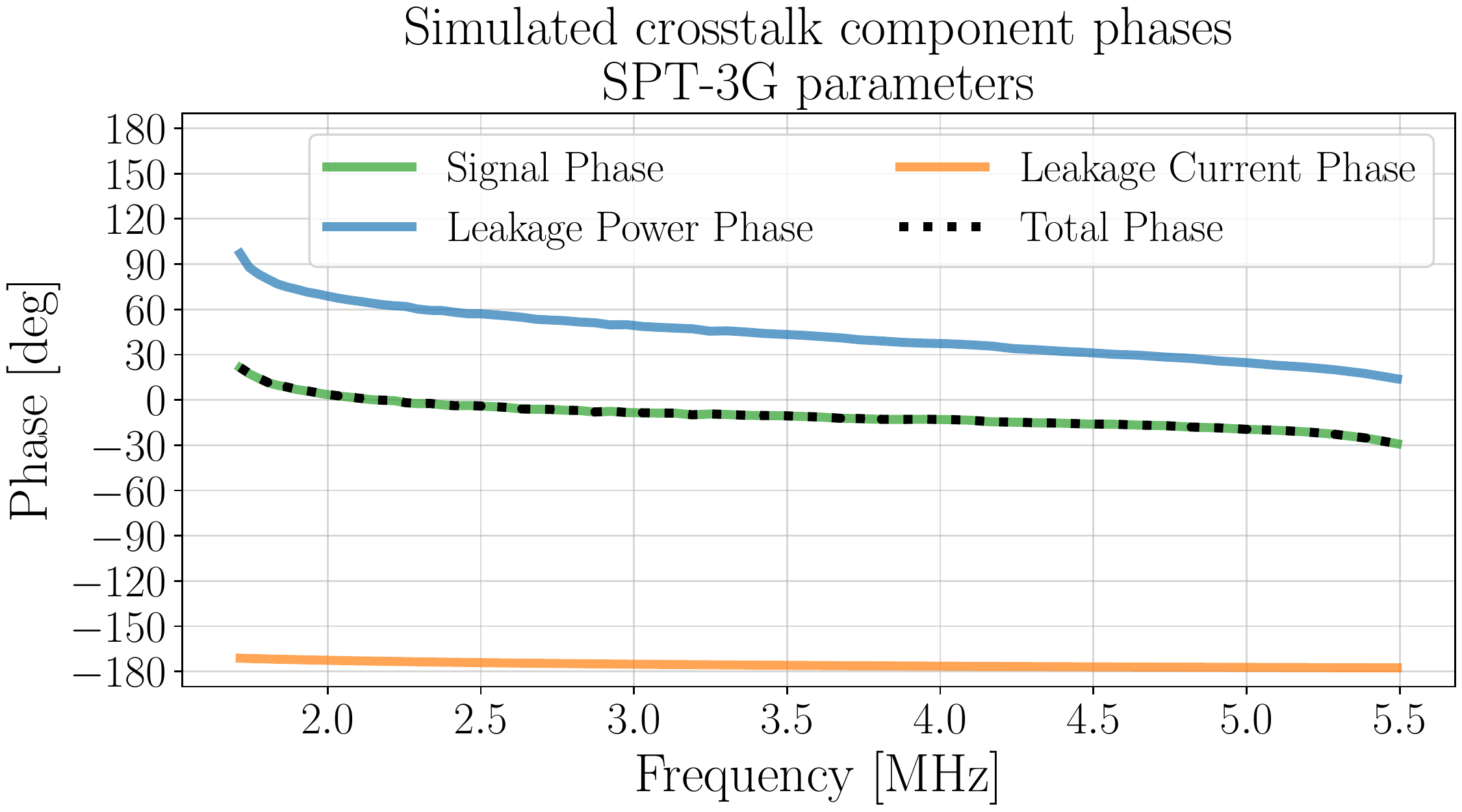}
	\end{subfigure}	
	\par\bigskip\bigskip
	\begin{subfigure}[b]{0.7\textwidth}
		\includegraphics[width=\textwidth]{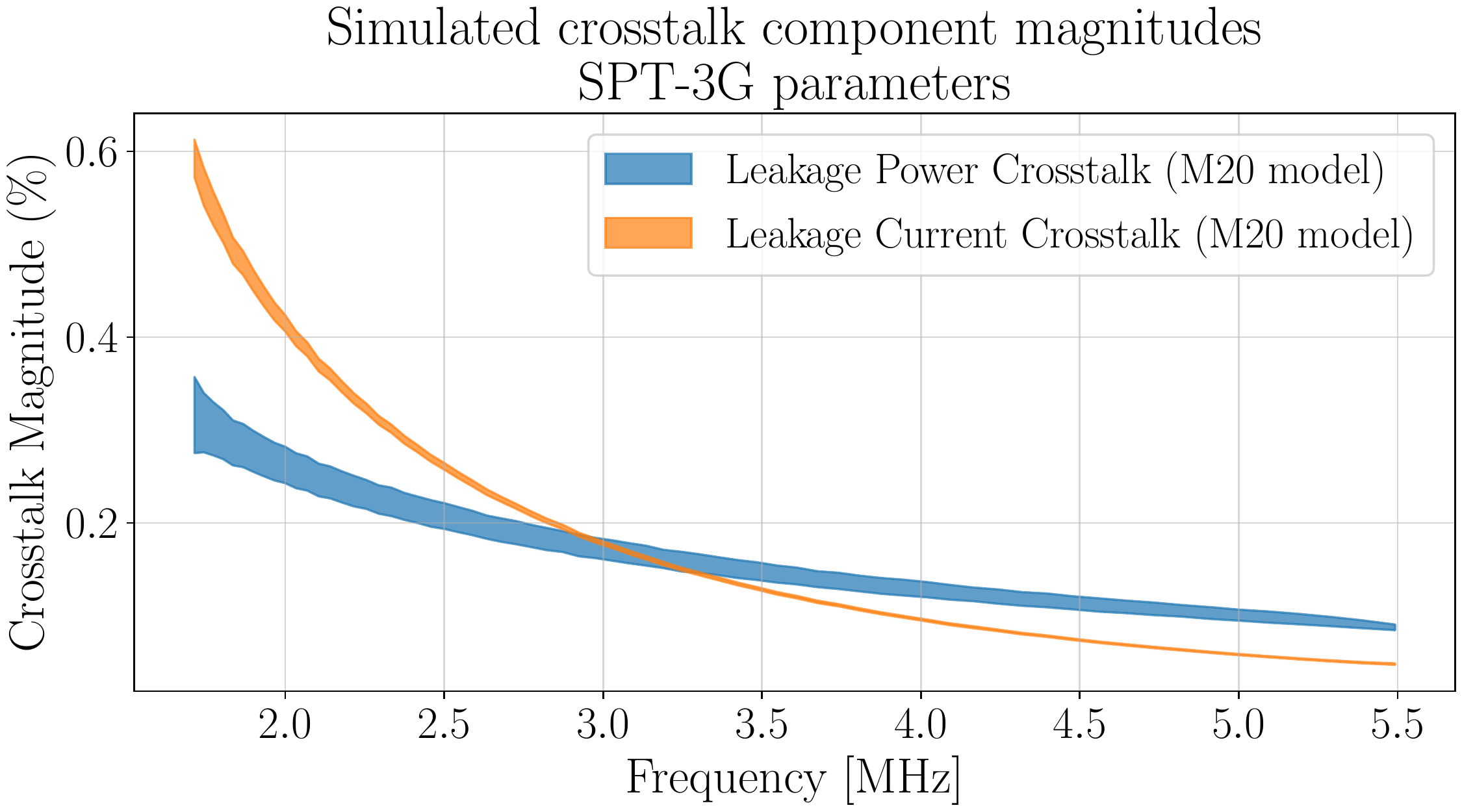}
	\end{subfigure}	 
	\vspace{0.5cm}	
	\captionsetup{width=0.8\linewidth} 
	\caption[Crosstalk component phase and magnitude.]{\textbf{Top:} The phase of each crosstalk component for a simulated \sptg{} multiplexing module with typical properties. The phases $\delta I_{i,\mathrm{signal}}$ and $\delta I_i$ are nearly identical because the total crosstalk fraction is very low. Most striking, the relative phase offset between the leakage power and the primary signal ranges from between approximately \ang{90} and \ang{45}, suppressing the effective contribution of leakage power crosstalk. \textbf{Bottom:} The magnitudes of each crosstalk component for an \sptg{}-like system. Line widths indicate the difference between ($i,i-1$) and ($i,i+1$) nearest-neighbor pairs. As magnitudes, both crosstalk quantities are shown as positive; but the relevant quantity is the vector-sum of both crosstalk vectors that is in-phase with the signal. Because leakage current crosstalk is approximately 180 degrees out-of-phase with the signal phase it appears as a negative crosstalk. Leakage power crosstalk is, for the most part, less than 90 degrees out-of-phase with the signal and appears as positive crosstalk. These two end up partially canceling.}
	\label{fig:crosstalk_phase}
\end{figure}

\begin{figure}[hbtp]
	\centering
	\includegraphics[width=0.8\linewidth]{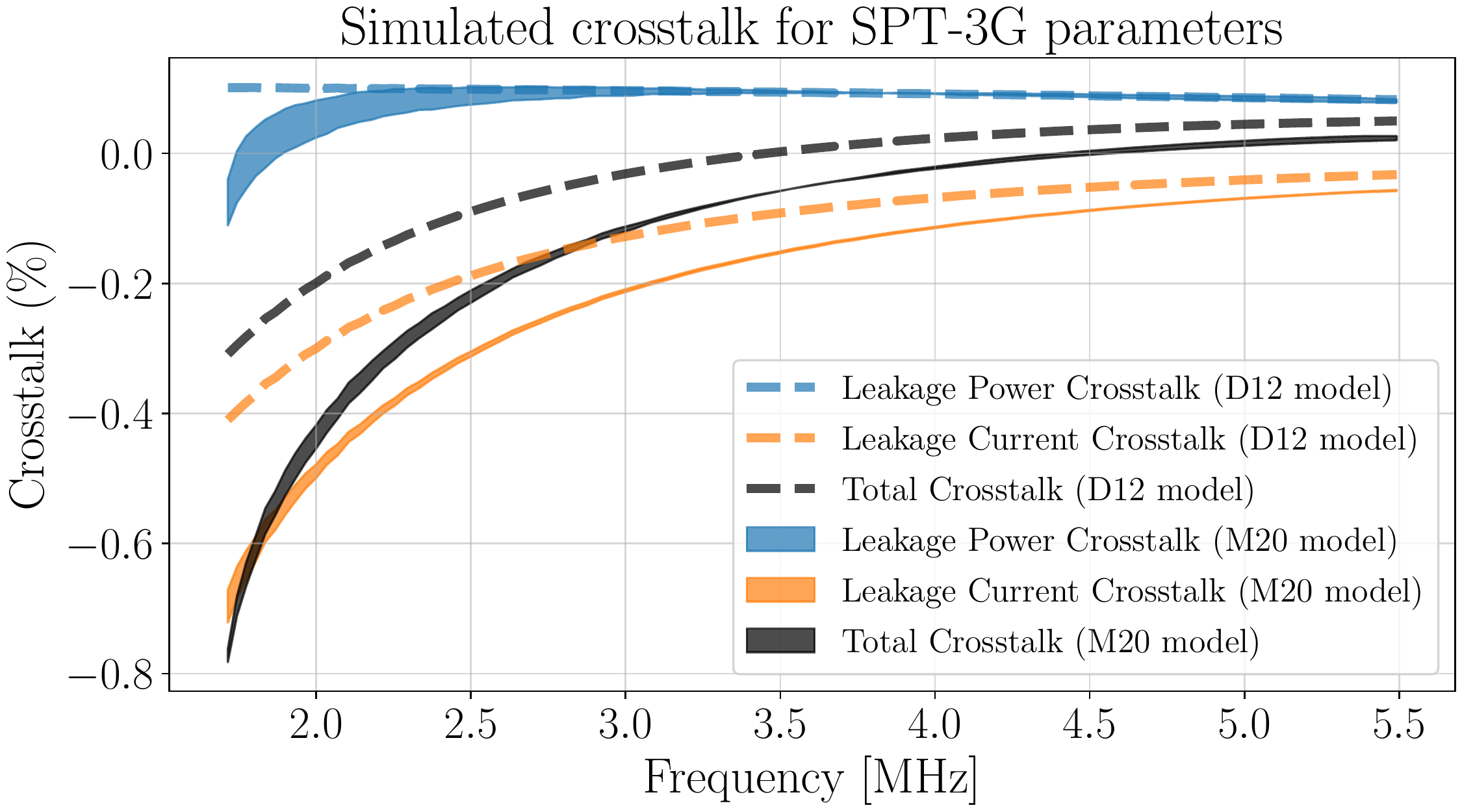}
	\vspace*{0.4cm}	
	\captionsetup{width=0.8\linewidth}
	\caption[Effective crosstalk.]{A comparison of the effective crosstalk fractions for a multiplexing module with typical \sptg{} parameters. Each mechanism is calculated via the derivations provided here and from \citetalias{dobbs12b}. The stray series resistance $r_s$ is responsible for a larger leakage current contribution than calculated using the \citetalias{dobbs12b} derivations. The dip in leakage power crosstalk at low frequency is due to the increased suppression from a $\sim$\ang{90} phase offset relative to the primary signal. The \citetalias{dobbs12b} model is largely insensitive to differences between ($i,i-1$) and ($i,i+1$) nearest-neighbor pairs and so has no visible changes in line width.}
	\label{fig:crosstalk_compare}
\end{figure}

The expressions for each signal described above have significant (and different) imaginary components, indicating they are all shifted in phase with respect to one another. 
All DfMUX systems record the complex signature of $I_{i}$, and the phase of the primary signal $\delta I_{i,\mathrm{signal}}$ is separately measured in-situ as part of the calibration for each observation.\footnote{More precisely, the phase of $\dfrac{\delta I_i}{\delta P_\nu}$ is measured in-situ, as radiative loading on the focal plane ($P_\nu$) is varied. Because the total crosstalk fraction is $<$1\% this is a good approximation of the phase of $\delta I_{i,\mathrm{signal}}$.} The final data product in the time domain is the projection of $I_{i}$ in-phase with the primary signal $\delta I_{i,\mathrm{signal}}$ \cite{montgomery20}.
The total crosstalk fraction is therefore the vector sum of two out-of-phase copies of the crosstalk signal, which partially cancel and is then further suppressed as it is projected into the primary signal axis,
\begin{align}
	    \frac{\delta I_{i,n,\mathrm{xtot}}}{\delta I_{i,\mathrm{signal}}}
	    =
	    \frac{\boldsymbol{\delta I_{i,n,\mathrm{xtot}}}\cdot\boldsymbol{\delta \hat{I}_{i,\mathrm{signal}}}}{\boldsymbol{\delta I_{i,\mathrm{signal}}}}
	    =
	    \frac{(\boldsymbol{\delta I_{i,n,\mathrm{LC}}}+\boldsymbol{\delta I_{i,n,\mathrm{LP}}})\cdot {\frac {\boldsymbol{\delta I_{i,\mathrm{signal}}} }{\left\|\boldsymbol{\delta I_{i,\mathrm{signal}}} \right\|}}}{
	    \boldsymbol{\delta I_{i,\mathrm{signal}}}}\,
\end{align}
where the leftmost quantity is a scalar crosstalk fraction, the operator $\cdot$ denotes a dot product, and $\boldsymbol{\delta \hat{I}_{i,\mathrm{signal}}}$ denotes a unit vector in the direction of the vector $\boldsymbol{\delta I_{i,\mathrm{signal}}}$.
This is shown in Figure \ref{fig:crosstalk_phase}, where the phase (top) and magnitude (bottom) of each signal is shown; and in Figure \ref{fig:crosstalk_compare}, where the effective crosstalk along the primary signal axis is plotted. 
Figure \ref{fig:crosstalk_compare} shows how the \citetalias{dobbs12b} model underestimates leakage current crosstalk (due to the contribution from $r_s$) and overestimates leakage power crosstalk (at low frequencies, due to suppression when projected along the primary signal axis). In all figures, the ``M20 model'' refers to the set of equations derived in this section, which were first published in \citep{montgomery20b}.

The crosstalk cancellation may seem non-intuitive, but it makes sense qualitatively: a decrement in the \textit{n\textsuperscript{th}} TES resistance will generate an increment in the current at $\omega_{i}$ (due to leakage current crosstalk), but it will also increment the deposited leakage power across the \textit{i\textsuperscript{th}} TES, raising its resistance, and causing a decrement in current at $\omega_{i}$ (due to leakage power crosstalk). These forms of crosstalk therefore oppose one another, though not perfectly, due to a phase offset in their action.

\subsection{Crosstalk model validation}	\label{sec:xtalk_validation}

\begin{figure}[hbpt]
	\centering
	\includegraphics[width=0.7\textwidth]{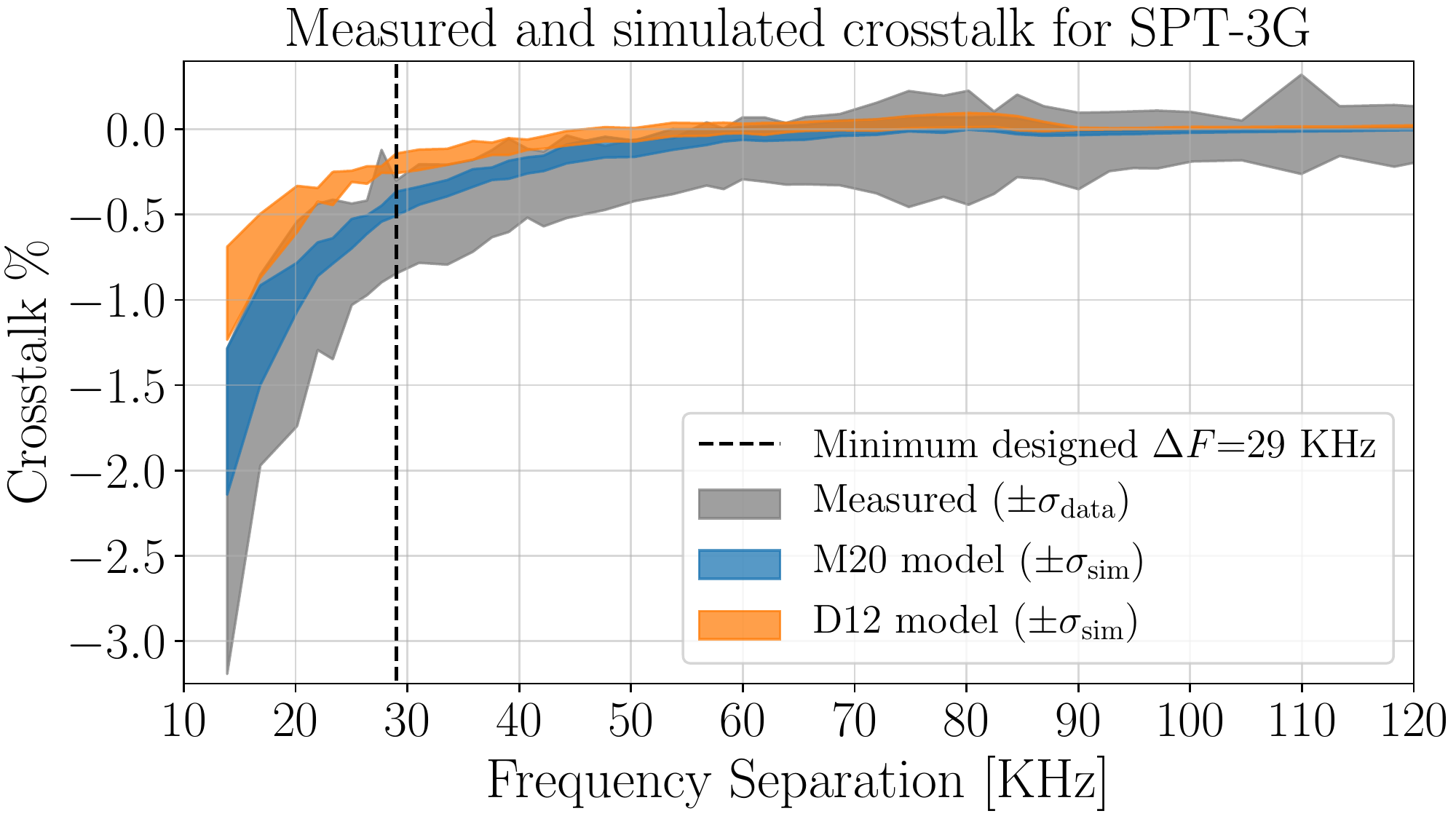}
	\vspace*{0.4cm}
	\captionsetup{width=0.8\linewidth}
	\caption[Comparison of measured and simulated crosstalk in \sptg{}.]{A comparison between measured in-situ crosstalk and simulated crosstalk for 4,400 detector pairs in the receiver. Each pair differs slightly in $r_s$, $R_\mathrm{TES}$, and underlying bias frequencies, all of which are measured separately \rev{and used to calculate the total expected crosstalk according to the M20 and \citetalias{dobbs12b} analytic models. These results, along with the measured crosstalk values for each detector pair, are shown binned by frequency separation with $\pm1\sigma$ of distribution in each bin.} The M20 model (blue), which includes stray impedances, is a better approximation for the measured crosstalk (gray) than the \citetalias{dobbs12b} model (orange), which omits stray impedances. The larger width in the distribution of measured crosstalk reflects the additional scatter associated with \rev{fitting noisy optical crosstalk measurements constructed from single-detector maps, and template-fitting of an extended source.}}
	\label{fig:sim_vs_measured_crosstalk}
\end{figure}

Measurements of crosstalk in-situ, using the extended source RCW38, have been previously described in \cite{bender19}. 
Comparison between these measurements and electrical crosstalk modeling is limited by the precision of the optical measurements, and the fact that such measurements can only be performed between detectors that have sufficiently different beams on the sky (so the crosstalk image can be distinguished from the source image).
The mapping between resonator frequency and detector on the focal plane is such that intended nearest-neighbors either observe the sky with overlapping beams (but orthogonal polarization), or with non-overlapping but adjacent beams (from a physically nearby pixel on the focal plane). This limits the above method to measuring crosstalk between detector-pairs designed to be non-nearest-neighbors. \rev{In most cases this results in large frequency separations and low crosstalk, but in rare cases the resonator frequencies have scattered due to variations in the fabrication, resulting in frequency spacing that is much narrower than intended.} In the latter case it is possible to measure nearest-neighbor-like electrical conditions between pixels that observe sufficiently separate regions of the sky.
Consequently, the best statistical test of the crosstalk is in the basis of bias frequency separation, rather than bias frequency, and a comparison in this basis between the measured optical crosstalk and the predicted crosstalk using the \citetalias{dobbs12b} and M20 models is shown in Figure \ref{fig:sim_vs_measured_crosstalk}. The region of low frequency separation allows an easy differentiation between the two crosstalk models.
Though not an ideal comparison, the result indicates the importance of stray impedances in crosstalk calculations, and supports the extension to analytic crosstalk modeling presented above.

\section{Readout noise modeling}\label{sec:noise}

Stray impedances also play an important role in noise performance. Instrument noise can be considered in two categories:
\begin{description}
	\item [Non-readout noise:] noise sources intrinsic to the TES operation and incident radiation. This category includes the photon noise (arrival time statistics of incident photons \cite{zmuidzinas03}) and phonon noise (random motion of thermal carriers that move heat away from the TES \cite{mather82}). These noise sources deposit power on the TES detector, which is converted to a current noise via the detector responsivity ($S$, in units of $\frac{\mathrm{pA}}{\mathrm{aW}}$).
	\item [Readout noise:] current noise sources that are additive with respect to non-readout noise, and independent of the detector responsivity. These are conventionally characterized as a noise equivalent current (NEI) spectral density in units of $\frac{\mathrm{pA}}{\sqrt{\mathrm{Hz}}}$ at the SQUID input. Sources of readout noise include: SQUID output noise, amplifier noise, transistor noise in the digital-to-analog converters (DACs) and analog-to-digital Converters (ADCs), Johnson-Nyquist noise from Ohmic elements in the signal path, and quantization noise from the digitization.\footnote{A detailed breakdown of these noise sources for \sptg{} can be found in \cite{montgomery20}.} For this discussion, TES Johnson noise is also included as a readout noise source, as it is suppressed when the TES responsivity is non-zero, but present when we measure readout noise \cite{lueker10b}.
\end{description}

Readout noise can be measured in-situ when detectors are in the normal state and thus have no responsivity to incident power ($S=0$), thereby disabling non-readout noise sources.
Although individual noise sources are almost identical in this generation of DfMUX to previous implementations \cite{dobbs12b}, NEI predictions based on models used for previous implementations poorly reconstruct the observed readout noise at higher bias frequencies. \rev{We identify two new mechanisms that modulate the existing noise sources and resolve this inconsistency. These effects occur due to parasitic impedances that only become relevant at higher frequencies, and are:}
\begin{description}
	\item[An effective low-pass output filter] between the SQUID output and room-temperature amplification stages. This filter attenuates signal and some, but not all, noise sources. A subset of readout noise sources in the output signal path are not attenuated, functionally amplifying that subset when referred back through the filter to the SQUID input. This output filter is described in more detail in Section \ref{sec:output_filter_tf}.
	
	\item[Parasitic capacitances to ground] in the cryogenic electronics, which generate a ``current sharing'' effect like the one first characterized in \cite{bender18}. In \cite{bender18}, a mechanism was identified by which nulling currents could avoid the SQUID input by flowing through the cryogenic filter network instead.
	Here we identify a second current path that bypasses the SQUID input, this time via parasitic capacitances to ground throughout the readout system. Similar to the output filter above, this results in an effective amplification of noise sources between the SQUID and ADC; but, unlike the output filter it also applies to the intrinsic SQUID noise, one of the largest sources of readout noise. It's possible to analytically calculate the resulting noise increase based on the cryogenic electronics design, as described in Section \ref{sec:current_sharing_tf}. Another consequence of this current path is that it partially spoils the differential balancing of the transmission lines going into the cryostat, making them more susceptible to radio frequency interference (RFI).
\end{description}

\begin{figure}[h!]
	\centering
	\includegraphics[width=0.8\textwidth]{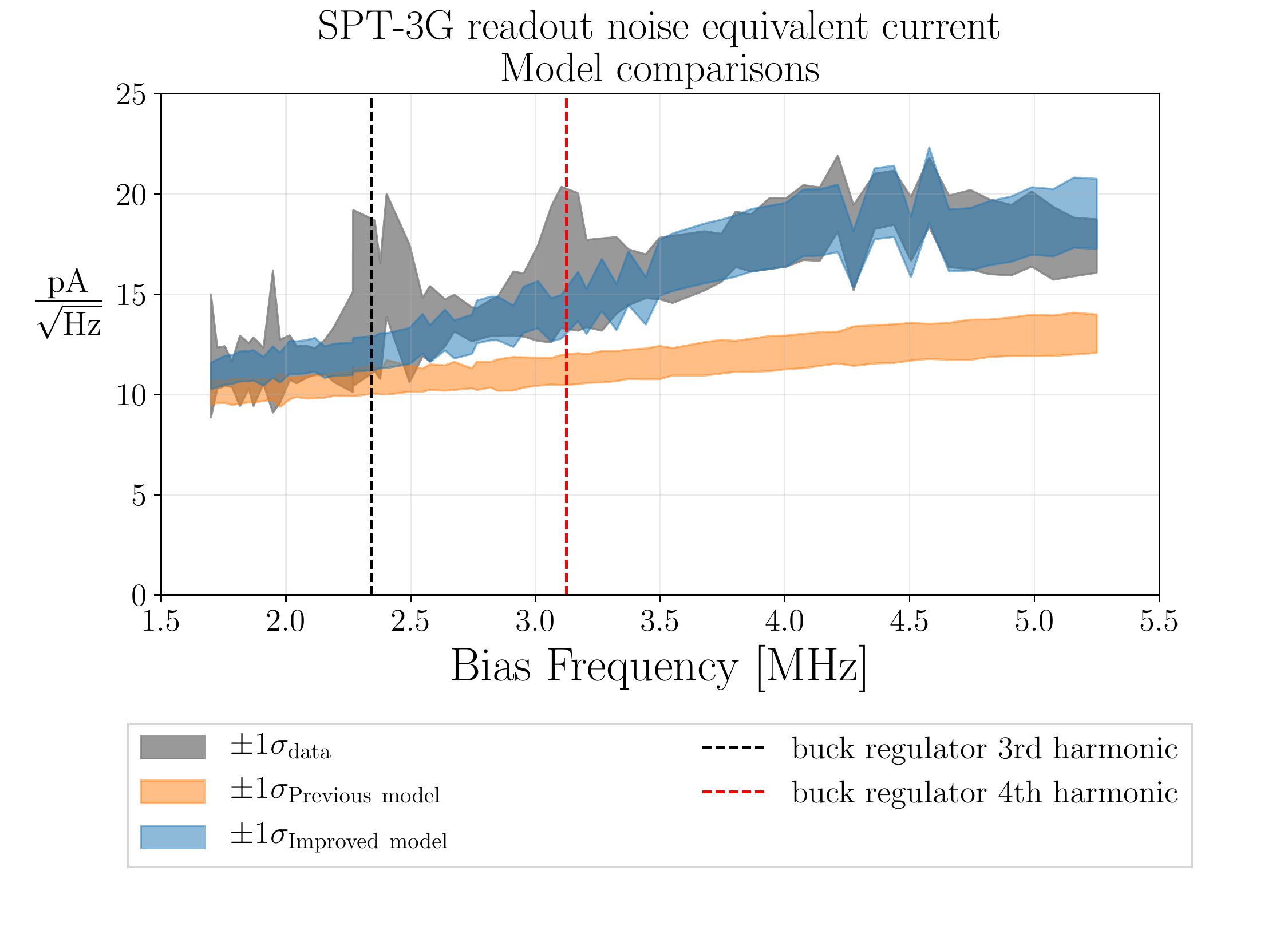}
	\captionsetup{width=0.9\linewidth}
	\caption[Improved noise model comparison.]{A comparison of measured \sptg{} readout noise with the previous noise model (orange) and an updated model (blue). The updated model includes the effects of the output filter and parasitic capacitances relevant to current sharing. The width of the distributions are calculated from the standard deviation of noise data and realizations binned by resonator (bias frequency). Noise realizations are generated using SQUID parameters corresponding to measured properties of SQUIDs in \sptg{}. Variation in SQUID performance is responsible for most of the scatter between detectors operated at the same bias frequencies. Dotted lines indicate the approximate switching frequencies of the buck regulators used to generate power in the warm electronics, and are a known source of additional noise not specifically captured in the noise model. Discontinuous steps as a function of bias frequency, most prominent at \SI{4.5}{\mega\hertz}, are due to changes in the geometric properties of the lithographic cryogenic filters for each resonator \cite{montgomery20}, and captured in the updated circuit model.}
	\label{fig:noise_model_comparisons}
\end{figure}

A full circuit model that includes both of these effects forms the core of an updated DfMUX noise model, \rev{which uses the software \texttt{PySPICE} \cite{pyspice} to numerically calculate some transfer functions based on the known circuit \cite{montgomery20}}. Figure \ref{fig:noise_model_comparisons} shows the measured readout noise alongside the previous and updated model expectations.
Besides describing the observed \sptg{} readout noise, these results suggest methods for improving noise performance in \sptg{} or future DfMUX readout designs, such as for \litebird{}.

\section{Output filter} \label{sec:output_filter_tf}	

\begin{figure}[hbpt]
		\centering
		\includegraphics[width=0.9\linewidth]{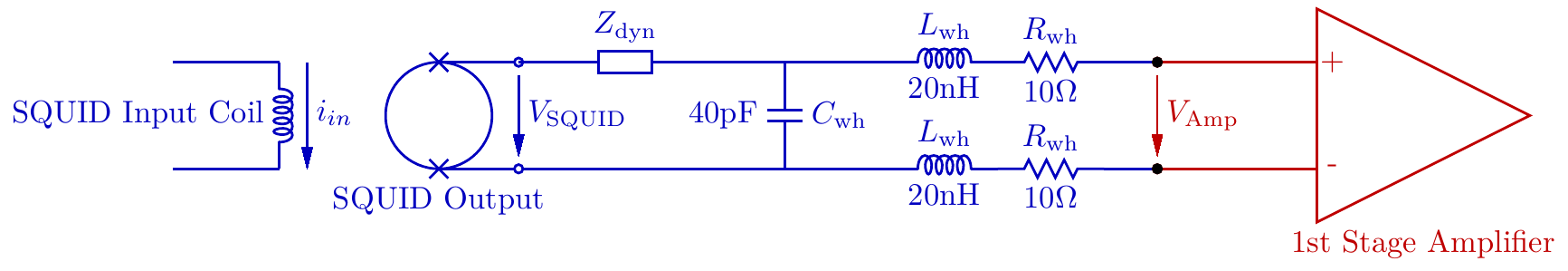}
		\vspace*{0.25cm}
		\captionsetup{width=0.95\linewidth}
		\caption[Demodulation signal path circuit diagram.]{The cryogenic portion of the output signal path is shown here in blue. The parallel capacitance within the wire harness ($C_\mathrm{wh}$) and SQUID dynamic impedance ($Z_\mathrm{dyn}$) form a low-pass filter that attenuates high frequency signals before they reach the 1st stage amplifier input.}
		\label{fig:output_filt_fig}
	\end{figure}

The signal path relevant to the output filter is between the SQUID output and the first room-temperature amplification stage, shown in Figure \ref{fig:output_filt_fig}.
The SQUID dynamic impedance ($Z_\mathrm{dyn}$) together with a parallel capacitance in the wire harness ($C_\mathrm{wh}$) generates a low-pass filter with a cutoff frequency $f_c\sim\frac{1}{2\pi Z_\mathrm{dyn} C_\mathrm{wh}}$. The SQUID dynamic impedance characterizes the relationship between the SQUID output voltage and current through the Josephson junctions that form the SQUID output, \cite{clarke04}
\begin{equation}
Z_\mathrm{dyn} = \frac{\delta V_\mathrm{out}}{\delta I_\mathrm{jj}} \,.
\end{equation} 
It is a SQUID property that is easily measured in-situ, and can be modeled as a real resistance in series with the voltage output signal.
The transfer function describing the resulting attenuation of the voltage signals is defined as $\chi_\mathrm{output} = \frac{V_\mathrm{Amp}}{V_\mathrm{SQUID}}$.

\begin{figure}[hbpt]
	\centering
	\includegraphics[width=0.6\linewidth]{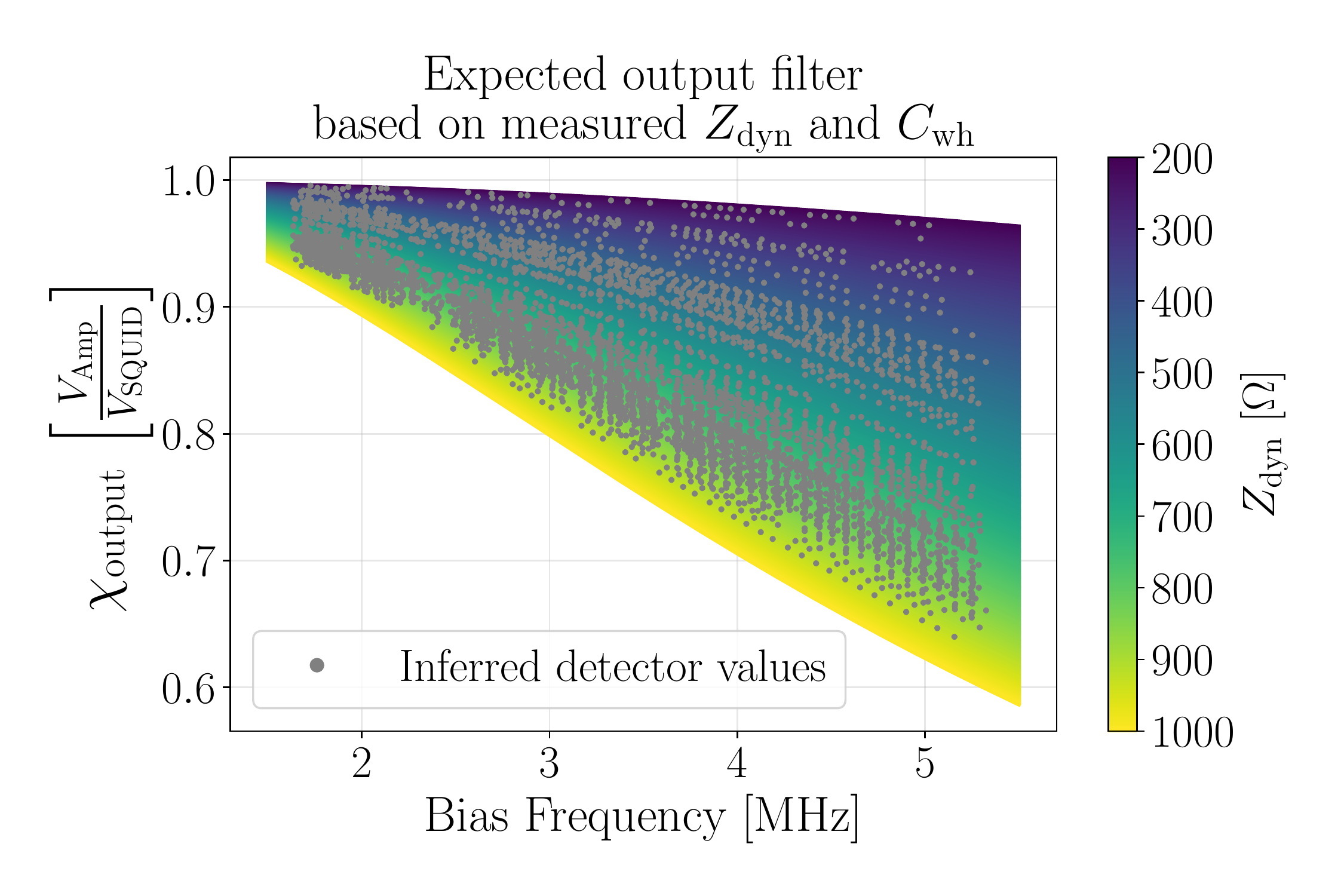}
	\captionsetup{width=0.7\linewidth}
	\caption[Measurements of $\chi_\mathrm{output}$ in \sptg{}.]{An analytic calculation of the output filter ($\chi_\mathrm{output}$) as a function of SQUID dynamic impedance ($Z_\mathrm{dyn}$) for the measured value of $C_\mathrm{wh}=\SI{40}{\pico\farad}$. Grey points indicate inferred values for each detector in the receiver, showing that although some detectors benefit from low dynamic impedance SQUIDs, most detectors are significantly attenuated by this filter, especially at high bias frequency. Noise sources in the output signal path are referred to an NEI at the SQUID input by dividing by $\chi_\mathrm{output}$, and therefore appear amplified by this filter.}
	\label{fig:chi_out_receiver}
\end{figure}

The value of approximately \SI{40}{\pico\farad} for $C_\mathrm{wh}$ is determined empirically, and is consistent with theoretical values for the wire harness design of approximately \SI{18}{\centi\meter} of 38 AWG Manganin twisted pair. 
\sptg{} SQUIDs exhibit a median dynamic impedance of \SI{750}{\ohm}, although there are six outliers that are operated in a low dynamic impedance configuration of \SI{350}{\ohm}.
Figure \ref{fig:chi_out_receiver} shows the resulting $\chi_\mathrm{output}$ for each detector in the \sptg{} receiver, inferred based on measurements of  $C_\mathrm{wh}$ and  $Z_\mathrm{dyn}$.

\begin{figure}[hbpt]
	\centering
	\includegraphics[width=0.8\linewidth]{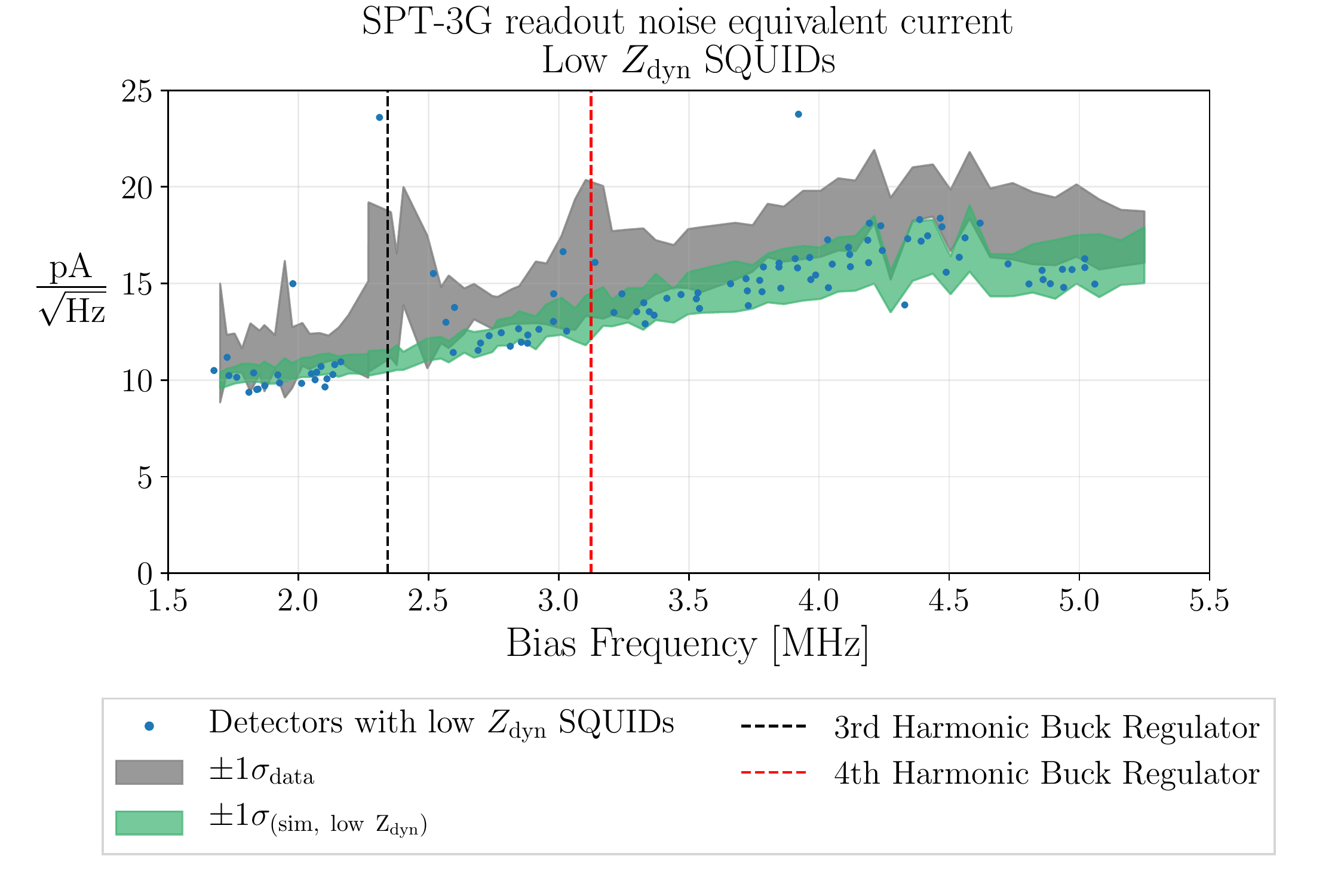}
	\captionsetup{width=0.9\linewidth}
	\caption[\sptg{} readout noise with low $Z_\mathrm{dyn}$ SQUIDs.]{Measured readout noise in detectors operated with SQUIDs exhibiting a lower dynamic impedance (blue points above) is systematically lower than the rest of the receiver (shown in gray as in Figure \ref{fig:noise_model_comparisons}). This difference is consistent with the expectation from the noise model, which predicts a $\sim$10\% noise improvement at high frequency from lower dynamic impedance SQUIDs. This model is shown in green, where low $Z_\mathrm{dyn}$ operation of the entire receiver is simulated. There are currently 6 SQUIDs on \sptg{} operated at low $Z_\mathrm{dyn}$.}
	\label{fig:lowz_noise}
\end{figure}

The six SQUIDs in \sptg{} operated with low dynamic impedance allow us to verify this model and the impact on readout noise performance. Figure \ref{fig:lowz_noise} shows measured readout noise for detectors associated with the low dynamic impedance SQUIDs. These detectors exhibit significantly lower noise than the receiver distribution, and are consistent with an expectation generated from noise realizations in which all SQUIDs exhibit similar \SI{\sim350}{\ohm} dynamic impedance.

In principle, all SQUIDs in the \sptg{} receiver can be operated with lower dynamic impedance; however, in nearly all cases this generates pathologies in the SQUID response that make them unusable. This appears to be related to the change in output filter cutoff frequency, which fails to attenuate out-of-band signals as strongly. Resonances in the SQUID design make them susceptible to such out-of-band signals, which then couple back into the SQUID input \cite{montgomery20}. 
In laboratory tests, \sptg{} SQUIDs become well-behaved when operating at low dynamic impedance if the stronger output filter is reproduced by artificially increasing $C_\mathrm{wh}$.
It's not clear why the six SQUIDs operated at low dynamic impedance in \sptg{} do not exhibit degraded performance in this configuration, as they are otherwise of the same design. 
The potential improvement to \sptg{} to developing a way to operate the entire receiver in a low dynamic impedance configuration is minimal relative to the risks associated with any modifications. However, this effect is an important consideration in the design and requirements for the SQUIDs and wire harnesses of the \litebird{} space telescope.

\section{Current sharing} \label{sec:current_sharing_tf}

Recall from Section \ref{sec:nulling} that Digital Active Nulling uses feedback in discrete bandwidths, centered at the carrier frequencies, to minimize signals at the digital demodulation stage (labeled ``DEMOD'' in Figure \ref{fig:dfmux_simple}, and digitized at the ADC). DAN injects whatever current at the SQUID input is necessary to minimize signals in this bandwidth.
In most cases this is equivalent to minimizing current through the SQUID input itself, which is the intended outcome. In other cases it requires driving current through the SQUID input. These two cases can be summarized as:

\begin{enumerate}
	
	\item Signals sourced in the input circuit, which exist as physical currents at the SQUID input. This includes the carrier tone and associated side-bands, as well as detector noise sources and readout noise sources in the carrier and nuller signal paths. When DAN injects nulling current to cancel voltages at the ADC that come from these sources, it does so by perfectly matching the amplitude of the currents that physically exist at the SQUID input, thereby both canceling them and faithfully recording them.
	
	\item Signals sourced in the output circuit, such as noise generated in the output path between the SQUID and the ADC, produce voltages at the ADC but do not exist as physical currents at the input to the SQUID. To minimize these signals at the ADC, DAN will drive current through the SQUID to generate opposing voltages. 
\end{enumerate}

Nulling currents of the second variety, which do not cancel existing currents, have several parallel paths through which to return, and only one of these is through the SQUID input. 
A fraction of the delivered waveform bypasses the SQUID input by flowing through these other parallel paths. 
Since only current flowing through the SQUID input will cancel noise generated in the output path, DAN must produce a larger copy of that noise to compensate for the portion lost via the parallel paths.
Like with the output filter, this is equivalent to a transfer function effect that amplifies noise sources in the output path; unlike the output filter effect, this amplification applies to the intrinsic SQUID noise as well.

\begin{figure}[hbpt]
	\centering
	\includegraphics[width=0.75\linewidth]{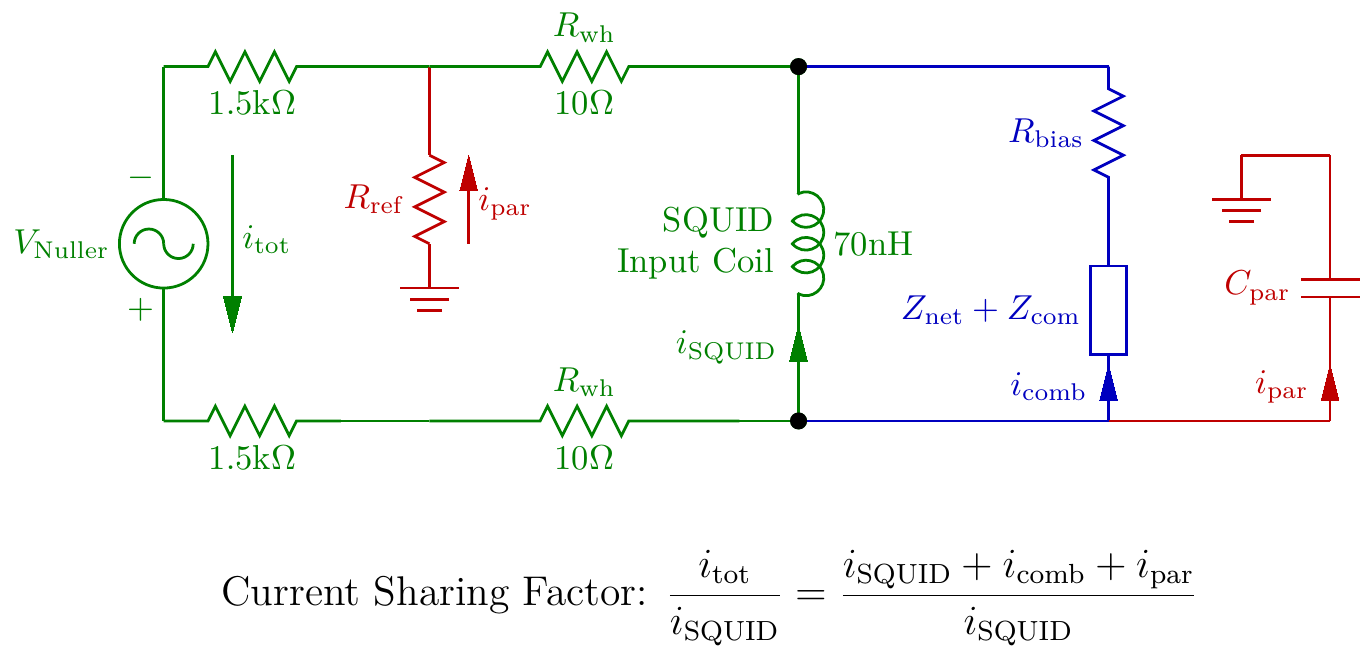}
	\vspace*{0.2cm}
	\captionsetup{width=0.9\linewidth}	
	\caption[Simplified current sharing circuit diagram.]{A simplified diagram based on Figure \ref{fig:circuit_model_complete}, highlighting the different parallel paths for an input nuller current. Signals that require DAN to drive a current through the SQUID input coil will be shared between three possible current paths, shown in green, blue, and red. The green path is the intended outcome, the blue path is an inevitable source of current sharing due to the topology of DfMUX systems, and the red path is due just to parasitic impedances in the system. Future receiver designs can mitigate this path with a re-engineering of a few electrical elements. The current sharing factor is the factor by which noise sources in the output path are amplified.}
	\label{fig:current_sharing_figure}
\end{figure}

The term given to the mechanism by which current is diverted around the SQUID input is current sharing, and the factor by which noise sources in the output path are amplified is the current sharing factor.
The three parallel paths for nuller currents are shown in a simplified schematic in Figure \ref{fig:current_sharing_figure} (based on Figure \ref{fig:circuit_model_complete}).
These are:
\begin{enumerate}
	\item Through the SQUID input coil ($|j\omega L_\mathrm{squid}|\approx 2.3\Omega$ at the highest bias frequencies) and back through the wire harness ($|R_{wh} + j\omega L_{wh}|\approx 10\Omega$), for a total of \order(\SI{10}{\ohm}). This is the desired current path, and ideally is the lowest impedance option to limit the current sharing factor in the system.
	\item Through the striplines ($Z_\mathrm{com}\approx 0.5\Omega$ at optimal bias frequencies), across the filtering network ($Z_\mathrm{net} \approx 1.7\Omega$), through the low impedance leg that generates the bias voltage ($|R_\mathrm{bias} + j\omega L_\mathrm{bias}|<<1\Omega$), and back out through the wire harness ($|R_{wh} + j\omega L_{wh}|\approx 10\Omega$), for a total of \order(\SI{10}{\ohm}). This path shares the final leg through the wire harness with (1). If this were the only other parallel path, the current sharing factor would be determined by a comparison between the SQUID input reactance and $Z_\mathrm{net}+Z_\mathrm{com}$, and come to a factor of approximately 1.7. This path is unavoidable for any DfMUX system, and its noise effects have been previously reported and accounted for.
	\item Through parasitic capacitances to ground within the signal chain, including within SQUID card wiring, lithographic filters, and TES wafer, and returning through $R_\mathrm{ref}=0~\Omega$. Together these impedances can be $\sim$\SI{20}{\ohm} at the highest bias frequencies, making it a significant path through the system, primarily because it avoids the $\sim$\SI{10}{\ohm} contribution from the wire harness. This is possible for two reasons: first, because the ground inside the cryostat is intentionally well-coupled to the ground outside the cryostat through structural and cryogenic elements; and second, because $R_\mathrm{ref}$ is low impedance. Once this path is included, the current sharing factor jumps to over 2.5. This path exists only due to stray impedances in the system and is not fundamental. Future receiver designs could mitigate this with a re-engineering of a few electrical elements.
\end{enumerate} 
Current sharing was first noted in \cite{bender19}, where the second path above was identified. We now additionally identify the third path through the parasitic capacitance to ground as a major contribution to a high current sharing factor in \sptg{}.

An exact calculation of the current sharing for each resonator requires a numerical simulation of the full circuit model, but the general form of the effect can be approximated by
\begin{align}  \label{eq:cs}
\chi_\mathrm{cs} &=\frac{I_\mathrm{SQUID~input~coil}}{I_\mathrm{Nuller~input}} \\ 
\begin{split}
&\approx
\biggl|\left(
\frac{
	((Z_\mathrm{com}(\omega)+Z_\mathrm{net}(\omega)+R_\mathrm{wh})\parallelsum j\omega L_\mathrm{SQUID})+ R_\mathrm{wh} + Z_\mathrm{parasitic}}
{Z_\mathrm{parasitic}}
\right) \\
&\qquad\times
\left( 
\frac{j\omega L_\mathrm{SQUID}+Z_\mathrm{com}(\omega)+Z_\mathrm{net}(\omega)}
{Z_\mathrm{com}(\omega_{i})+Z_\mathrm{net}(\omega)}
\right)
\biggr|
\,,
\end{split}
\end{align}
for a $Z_\mathrm{parasitic}$ that corresponds to the effective capacitive reactance of the parasitic current path.
The current sharing factor is then given by $1/\chi_\mathrm{cs}$.

\subsection{In-situ current sharing model validation}

\begin{figure}[hbpt]
	\centering
	\includegraphics[width=0.7\linewidth]{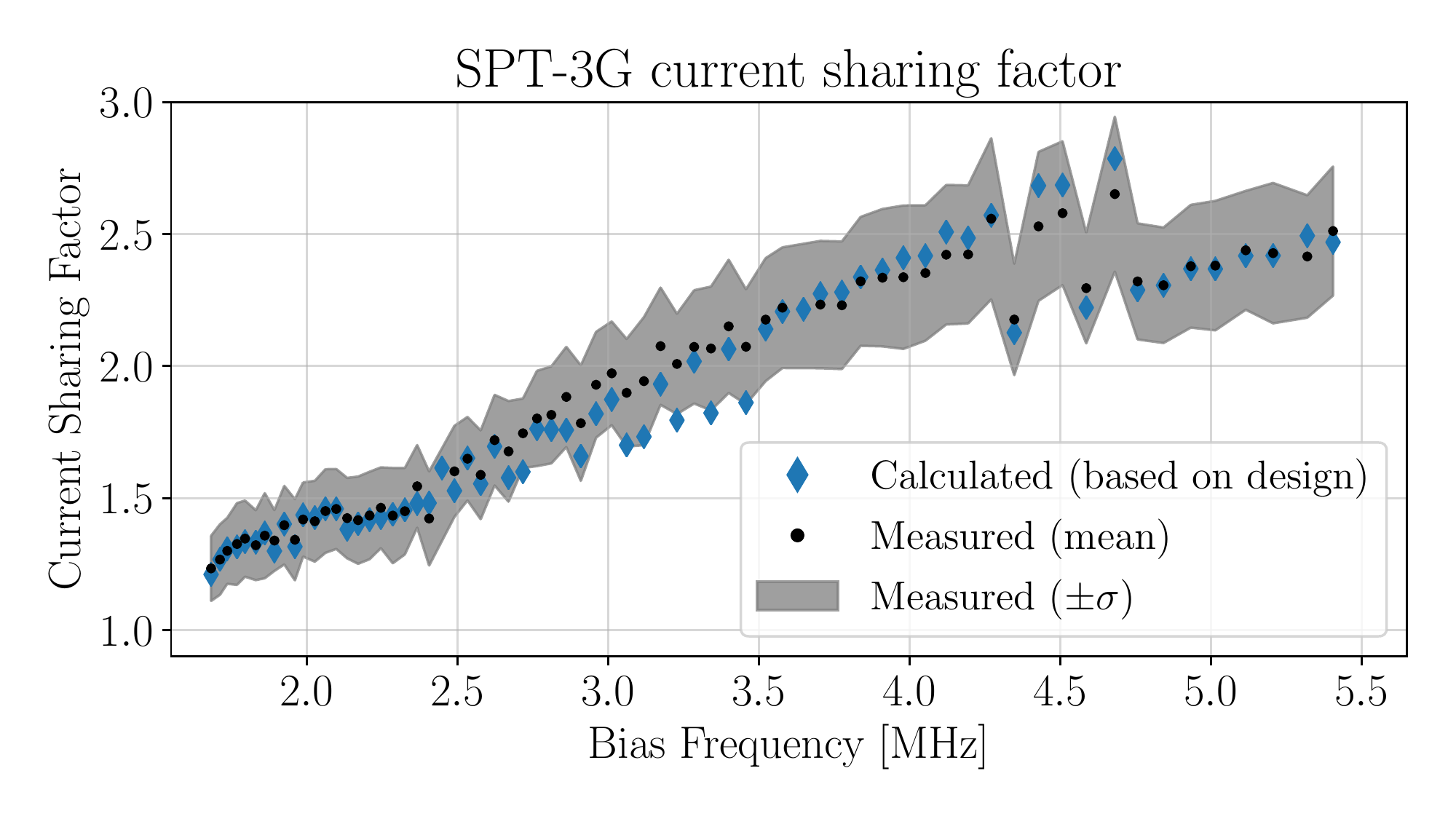}
	\captionsetup{width=0.9\linewidth}
	\caption[Measured and simulated current sharing factor.]{The current sharing factor $\left(1/\chi_\mathrm{cs}\right)$ can be measured directly in-situ and compared to an expectation based on an analytic evaluation of the electrical model shown in Figure \ref{fig:circuit_model_complete}, which includes both current sharing paths. The plot above compares the distribution of measured values for each detector across \sptg{} (mean and standard deviation, binned by resonator on the multiplexing module) with a fully analytic expectation based on the readout circuit design. The circuit simulation assumes designed values for all parameters, while the hardware in the receiver includes scatter in detector and filter properties, so width in the measured distribution is expected. The agreement between the simulations and measured values is excellent, and the largest disagreement (between \SIrange{3}{3.5}{\mega\hertz}) corresponds to a known region of high scatter in the fabricated LC resonant frequencies, such that neighboring channels are more likely to deviate from the circuit model and have overlapping filters \cite{montgomery20}. This suggests that an analytic model for the readout is accurately capturing the relevant dynamics for current sharing effects.}
	\label{fig:measured_cs_factor}
\end{figure}

A direct measurement of the current sharing factor is possible in-situ by comparing the amplitude of a known nuller input current to the signal measured at the SQUID output.
Figure \ref{fig:measured_cs_factor} shows the measured distribution of current sharing factor at each bias frequency in the \sptg{} system alongside a prediction based on the analytic circuit model.

\subsection{Experimental current sharing model validation}

\begin{figure}[htp]
		\centering
		\includegraphics[width=0.75\textwidth]{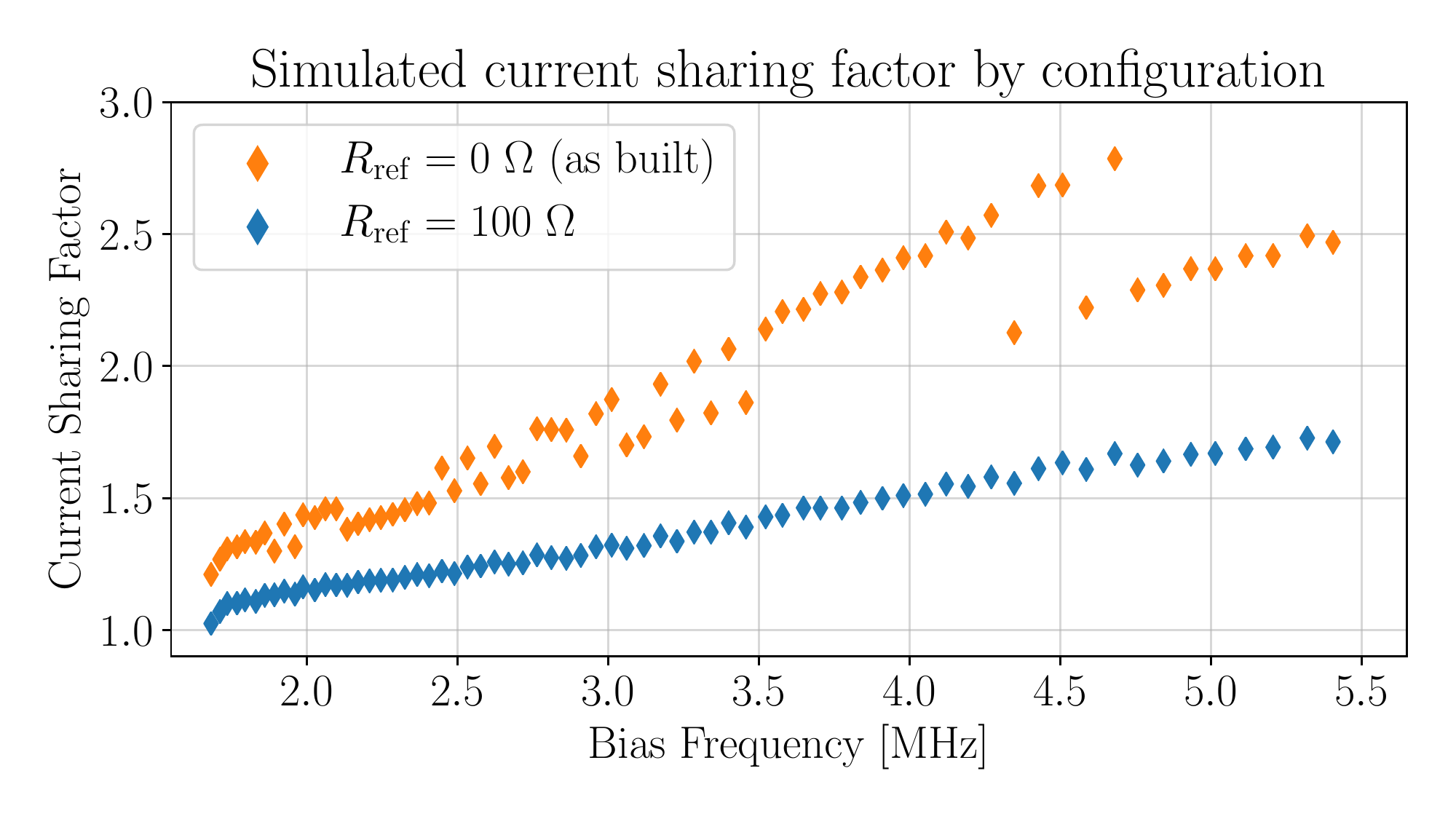}
		\captionsetup{width=\linewidth}
		\vspace*{-.25cm}
		\caption[Simulated current sharing factor with parasitic current modifications.]{The low impedance ground reference allows a current sharing path through parasitic capacitances in the system, resulting in a substantial current sharing factor. Modifying this reference to increase the resistance to \SI{100}{\ohm} disables that current sharing path and improves the current sharing factor, as simulated above. By reducing the current sharing factor, the noise performance of the system improves, as demonstrated in Figure \ref{fig:rmod_noise}.}
		\label{fig:cs_with_r48mod}

		\vspace*{1cm}
		\includegraphics[width=0.8\textwidth]{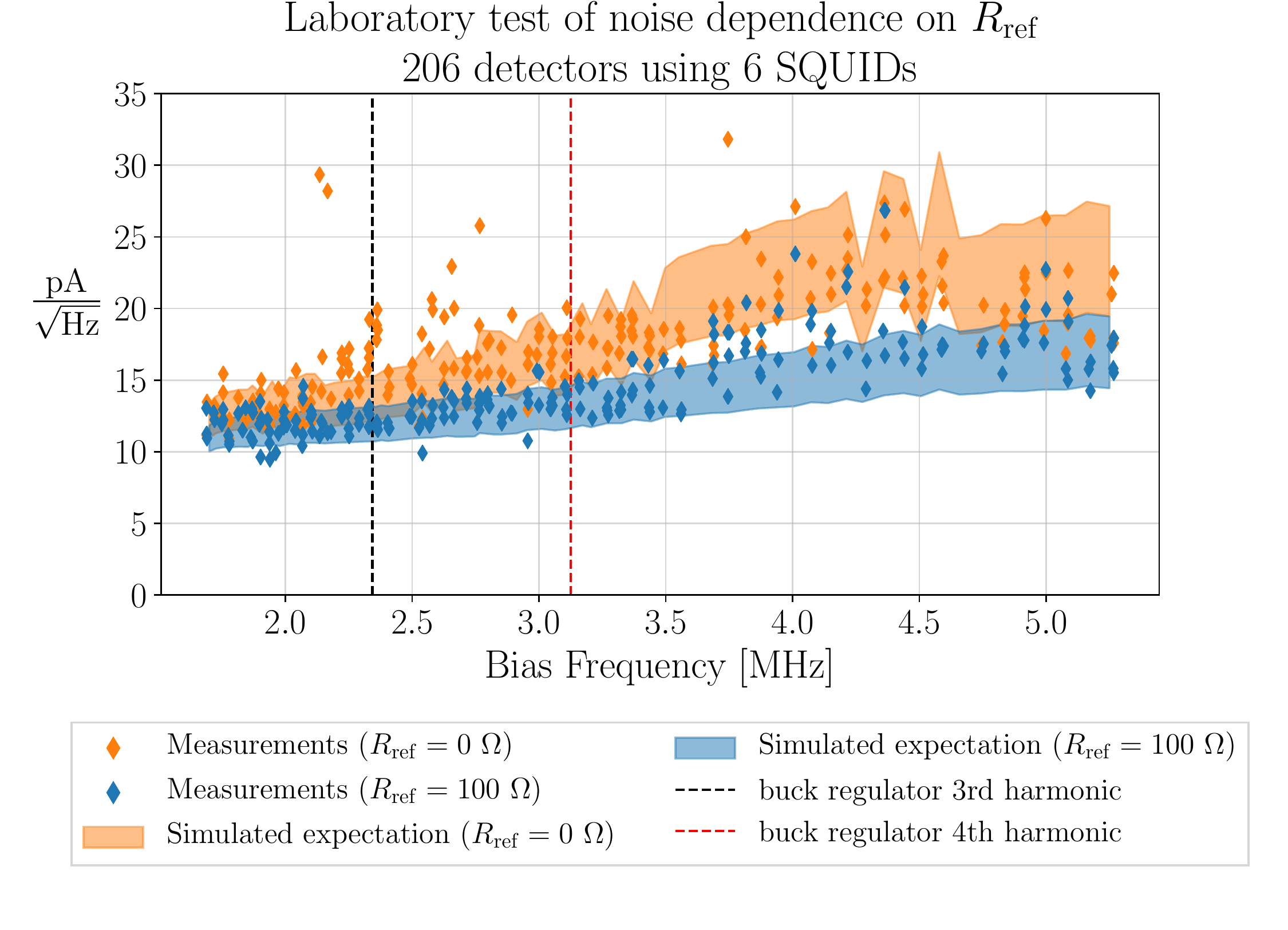}
		\captionsetup{width=\linewidth}
		\vspace*{-.25cm}		
		\caption[Measured readout noise with modification.]{Laboratory tests modifying $R_\mathrm{ref}$ confirm a substantial reduction in readout noise, consistent with a change in the current sharing factor as shown in Figure \ref{fig:cs_with_r48mod}. The highest frequency channels under-perform expectations, possibly indicating an additional current return point not yet characterized, though this may also be a feature of laboratory test environment. Separately, the modification improves scatter, particularly in channels near to the buck regulator frequencies, which suggests an improvement to the differential balancing and consequent electromagnetic inference susceptibility.}
		\label{fig:rmod_noise}
\end{figure}

Although there are many relevant capacitances to ground throughout the system, the return paths for all of them flow through $R_\mathrm{ref}$, which is currently a \SI{0}{\ohm} ground reference in the room-temperature electronics, provided by a single resistor. The nuller voltages ($V_\mathrm{Nuller}$) are transformer-coupled, and so a ground reference is required for the DC SQUID flux bias (shown in Figure \ref{fig:circuit_model_complete}) to produce a bias current through the SQUID input, but that reference is not required to be low impedance. That choice was intended to prevent any ambient electromagnetic interference from generating voltages at the input of the SQUID, but the intent is undermined by enabling such a significant current sharing mechanism.
Figure \ref{fig:cs_with_r48mod} shows the simulated improvement in current sharing factor with $R_\mathrm{ref}=\SI{100}{\ohm}$.
We validate this model (and potential improvement) using laboratory measurements of 6 SQUID modules, shown in Figure \ref{fig:rmod_noise}. The detectors and environment used for this test exhibit a higher characteristic readout noise than seen on \sptg{}, but the significance of the noise improvement after the modification to $R_\mathrm{ref}$ is large relative to that difference, and consistent with expectations based on the predicted reduction in current sharing. 

At the highest frequencies, the noise improvement is more modest than expected.
This may be a consequence of the small number of detectors, or a feature of the laboratory test setup. It could also be indicating an additional current return point that remains uncharacterized.
Nevertheless, this test demonstrates the utility of the readout model, which can reliably simulate changes in instrument performance as a function of design, and guide hardware changes to the electronics. The choice of \SI{100}{\ohm} has not been subject to optimization, and was selected to improve current sharing while keeping interpretation of the results simple. Larger-scale investigations of this are underway, and it has been incorporated into the baseline \litebird{} readout design \cite{montgomery20}.

\subsection{Modifying existing instruments}

Modifying $R_\mathrm{ref}$ on existing room-temperature readout electronics currently deployed on CMB instruments is relatively non-invasive, as it does not require access to the vacuum or cryogenic volumes of the instrument. Although such an intervention may take place for \sptg{} during the next austral summer maintenance period, any hardware modification carries intrinsic risk, and in this case that risk includes manipulation of electronics at the vacuum interface. While the above-shown experimental measurements indicate a possible $\sim$35\% improvement to the readout noise of detectors operated with the highest bias frequencies, the overall improvement to noise-equivalent temperature (including non-readout noise sources) is more modest, and any improvement to our target science analyses would be small.

\section{Conclusion}

To achieve multiplexing factors that enable modern CMB instruments, DfMUX readout systems now operate at megahertz bias frequencies with dense filter arrangements, and are exercising the electronics signal paths in ways that previous generations did not.
An extension of the analytic crosstalk and noise models is required for precision forecasting and analysis of these systems.
Presented here are new analytic formulas for crosstalk analysis, which better describe the \sptg{} performance, and allow for instrument designs to take advantage of the cancellation and projection effects in order to design a lower-crosstalk DfMUX system.
Also presented are a set of two previously unexplained mechanisms by which readout noise is amplified, which were responsible for discrepancies between prior noise models and measured noise performance. With this knowledge, these noise mechanisms can be mitigated in future implementations by making simple changes to the DfMUX circuit design.

\subsection*{Disclosures}
The authors have no relevant financial interests, or other potential conflicts of interest, in the manuscript to disclose. An earlier version of this manuscript appeared as an SPIE conference proceedings in 2020.

\subsection* {Acknowledgments}
The South Pole Telescope program is supported by the National Science Foundation (NSF) through grants PLR-1248097 and OPP-1852617.
Partial support is also provided by the NSF Physics Frontier Center grant PHY-1125897 to the Kavli Institute of Cosmological Physics at the University of Chicago, the Kavli Foundation, and the Gordon and Betty Moore Foundation through grant GBMF\#947 to the University of Chicago.
Argonne National Laboratory’s work was supported by the U.S. Department of Energy, Office of High Energy Physics, under contract DE-AC02-06CH11357. 
We also acknowledge support from the Argonne Center for  Nanoscale Materials.
Work at Fermi National Accelerator Laboratory, a DOE-OS, HEP User Facility managed by the Fermi Research Alliance, LLC, was supported under Contract No. DE-AC02-07CH11359.
The McGill authors acknowledge funding from the Natural Sciences and Engineering Research Council of Canada, Canadian Institute for Advanced Research, and the Fonds de recherche du Qu\'ebec Nature et technologies.
N. Whitehorn acknowledges support from NSF CAREER grant AST-0956135.
This manuscript was typeset using \LaTeX.
Many of the figures were made using \texttt{matplotlib} \cite{matplotlib} and PGF/Ti\textit{k}Z \cite{tikz}. Circuit diagrams were made using \texttt{PyCirkuit}, written by Orestes Mas.\footnote{\href{https://github.com/orestesmas/pycirkuit}{https://github.com/orestesmas/pycirkuit}}
Numerical circuit simulations were performed using \texttt{PySpice} \cite{pyspice}.
Analyses were conducted using \texttt{python} scientific packages \cite{scipy, numpy, ipython}.


\bibliography{../../BIBTEX/spt}   
\bibliographystyle{spiejour}   


\vspace{2ex}\noindent\textbf{Joshua Montgomery} is a postdoctoral researcher at McGill University. He received his BA from the University of Chicago in 2011, and MSc and PhD from McGill University in 2015 and 2020, respectively. He is a member of the South Pole Telescope and POLARBEAR collaborations, and wintered over at the South Pole as an operator for SPT-3G in 2018. He is currently a scientist and project manager for LiteBIRD Canada.

\vspace{1ex}
\noindent Biographies of the other authors are not available.

\listoffigures
\listoftables

\end{spacing}
\end{document}